\documentclass[aps,preprintnumbers,nofootinbib]{revtex4}
\usepackage{amsfonts,amssymb,amsmath}            
\usepackage[]{graphics,graphicx,epsfig}            
\usepackage{amsthm}
\usepackage{stmaryrd}                           
\def\identity{\leavevmode\hbox{\small1\kern-3.8pt\normalsize1}}
\theoremstyle{plain}
\newcommand{\ket}[1]{\left | \, #1 \right\rangle}

\input{Qcircuit}

\begin{document}

\title{Deriving a Fault-Tolerant Threshold for a Global Control Scheme}

\author{Alastair \surname{Kay}}
\email[]{a.s.kay@damtp.cam.ac.uk} 
\affiliation{Centre for Quantum Computation,
             DAMTP,
             Centre for Mathematical Sciences,
             University of Cambridge,
             Wilberforce Road,
             Cambridge CB3 0WA, UK}

\begin{abstract}
In this paper, adapted from the author's PhD thesis, we present otherwise unpublished results relating to global control schemes, culminating in the calculation of a fault-tolerant threshold for one such scheme. As with early fault-tolerant threshold results, the aim is to calculate a positive number, not to optimise it. We also discuss how the results might affect other related schemes, such as those based on cellular automata.
\end{abstract}

\maketitle

In some physics settings, such as optical lattices, while we can initialise states of a large number of qubits, the control of individual qubits is particularly challenging. In other systems, while single-qubit addressing may be feasible, the scaling requirements of needing many control elements for each and every qubit, and for the interactions between qubits, make device structures very complex, and it would be desirable to reduce these. In these realisations, we should see if an architecture can be designed that is suited to the physical situation, ensuring that it is at least as powerful as the theoretical architecture for Quantum Computation, by demonstrating a universal set of quantum gates.

In order to avoid single-qubit addressing, we assume that we have a set of fields that we can control. These fields address the whole device in some way, and it is our task to see how to compose these to give universal computation, and to address the parallelism requirements for error correction and fault-tolerance. In the early sections, we present a review of of the global control scheme we use \cite{Benjamin:2002a}, including the required structures for error correction and fault-tolerance of the computational qubits \cite{Benjamin:2003b} and the auxiliary (classical) states \cite{Kay:2005c}. We then detail the calculation of a fault-tolerant threshold in this restricted scenario, comparing the trade-offs of different assumptions. This is adapted from \cite{Kay:thesis}.

\section{Introduction to Global Control}\label{chap:5}

\subsection{State Transfer by Global Control} \label{sec:1}

\begin{figure}[t]
  \begin{center}
    \leavevmode
\resizebox{0.4\textwidth}{!}{\includegraphics{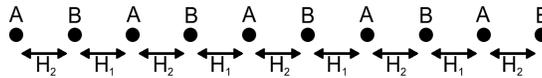}}
\end{center}
\caption[State transfer by global control]{We can transfer a quantum state from one end of the chain to the other end, simply by alternately applying the Hamiltonians $H_1$ and $H_2$.}
\label{fig:chain_definition}
\end{figure}

A relatively simple protocol that can be used to demonstrate some of the ideas of global control is that of state transfer. The scenario is that we start with a chain of qubits (Fig.~\ref{fig:chain_definition}), and the qubit at one end has some quantum state, $\ket{\psi}$, stored on it. We would like to transfer this state to the opposite end of the chain\footnote{We know how to do this with no time-varying interaction by engineering a fixed interaction Hamiltonian, as in \cite{Christandl,Kay:2004c,kay:2006a}, but let us disregard this for the moment.}. A typical way to do this would be to to perform {\sc swap} gates one at a time so that the state is discretely moved from qubit 1 to qubit 2, and then from qubit 2 to qubit 3 and so on. However, we can do exactly the same with global control. In this situation, we only allow pulses to be sent to the entire device. For example, we could allow a global pulse to turn on an interaction between alternate pairs of qubits. Specifically, one might think of turning on an interaction of the form
$$
H_1=\sum_n(X_{2n}X_{2n+1}+Y_{2n}Y_{2n+1})
$$
which gives a series of {\sc swap} operations. By alternating this with a second global field that activates the interaction
$$
H_2=\sum_n(X_{2n-1}X_{2n}+Y_{2n-1}Y_{2n})
$$
you can quickly convince yourself that it is possible to achieve the desired state transfer. It is pair-wise interactions of this form that we will use to create a general quantum computation protocol with global fields. This state transfer protocol has two notable advantages over the permanently coupled spin chain known from studies of perfect state transfer \cite{Christandl,Kay:2004c,kay:2006a}. Firstly, we can perform the transfer whenever we want, without the complications of moving the state onto an ancillary device. Secondly, the transfer occurs independently of the states of the other qubits in the system; there are no controlled-phase gates applied during transfer.

\subsection{Quantum Computation} \label{sec:2}

We shall now examine, following \cite{Benjamin:2002a}, how to perform a Quantum Computation on a one-dimensional chain of qubits with two switchable fields. We discuss this as it is the minimal case, so more complex systems should always be able to use this system, generally with refinements available which can significantly decrease the overheads involved in this simple scenario \cite{Benjamin:04}. The concept was originally introduced by Lloyd \cite{Lloyd:1993a}, and, although his constructions were not minimal, have formed the basis of all subsequent schemes.

Let us take a simple chain of qubits and allow pair-wise interactions between them. These should be grouped like the Hamiltonians $H_1$ and $H_2$, but allow general two-qubit gates instead of a simple {\sc swap} gate (denoted by $\beta$ and $\alpha$ respectively). Further, we shall make a distinction between the odd-numbered qubits (denoted A) and the even-numbered ones (B), as shown in Fig.~\ref{fig:chain_definition}. The desire for a general 2-qubit interaction may seem like we're demanding a lot of the system, but there are simple ways to rephrase this requirement. For example, we can consider the ability to perform arbitrary single-qubit rotations on all the A qubits or on all the B qubits. Coupling this with a single 2-qubit interaction, such as a controlled-phase, is enough to create any arbitrary two-qubit interaction \cite{nielsen}.

The first thing to note is that by alternating applications of $H_1\equiv \beta^{\text{{\sc swap}}}$ and $H_2\equiv\alpha^{\text{{\sc swap}}}$, we can move the states on the A qubits independently of those on the B qubits (this is just state transfer again), assuming a chain of infinite length. The way that we plan to implement the computation is to place computational qubits only on the As. We somehow initialise one of the B qubits in the $\ket{1}$ state and take every other qubit to be, initially, in the $\ket{0}$ state. At this stage it might appear that we are avoiding single-qubit control by creating a single-qubit state at the start of the computation. However, this is easier than single-qubit gates in general because there are other properties that we can take advantage of, such as edge effects at the end of the chain. How we perform the initialisation will be determined by the physical system \cite{Vollbrecht,kay-2004-6,kay-2006-73,Zoller:1}, but for the moment we can assume that we have control over a single qubit at the end of the chain. Our state transfer protocol can therefore be used to move this unique state, known as the Control Unit ({\sc cu}), such that it is adjacent to any A qubit that we desire. At that point, we can perform a gate $\beta^{c-U}$, i.e.~a controlled-$U$ operation, which means that the operation $U$ is performed on the qubit to the right of the {\sc cu}. Everywhere else, the control qubit is in the $\ket{0}$ state, and so nothing happens. As a result, we have performed a one-qubit gate on a specific qubit using global pulses.

\begin{figure}[!t]
\begin{center}
\includegraphics[width=0.45\textwidth]{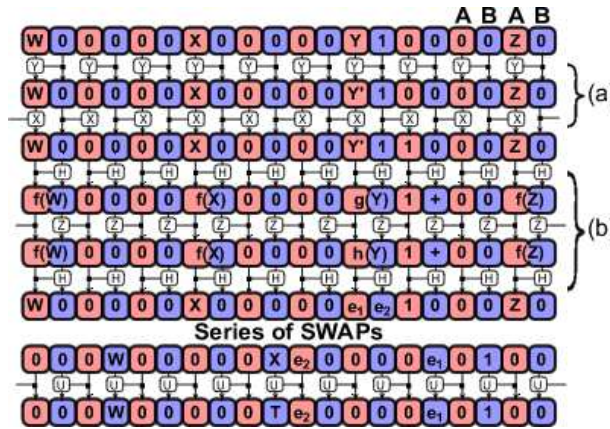}
\caption[Two-qubit gate by global control]{Method for creating a two-qubit gate between distant qubits. The sequence up to the {\sc swap}s places the information of the qubit Y onto the {\sc cu} ($\ket{e_1e_2}=\alpha\ket{10}-\beta\ket{01}$ where $\ket{Y}=\alpha\ket{0}+\beta\ket{1}$), and subsequently this is used to act upon the target qubit, X. After the controlled-$U$ step, all previous steps must be undone to disentangle the {\sc cu}. Part (a) translates the {\sc cu} into a unique local pattern in the $A$ qubits. Part (b) flips $B$ qubits if they are surrounded by two $\ket{1}$s. This is the mechanism that can be used to switch on and off sets of {\sc cu}s.}
\label{2_qubit_gate}
\end{center}
\end{figure}

\begin{figure}[t]
  \begin{center}
    \leavevmode
\resizebox{0.35\textwidth}{!}{\includegraphics{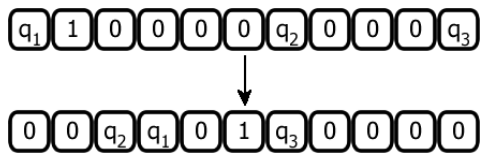}}
\end{center}
\caption[Minimal size of a globally controlled array]{The apparently minimal device size for three qubits. However, if we want to perform a one-qubit gate on the third qubit, we get a two-qubit gate between the other two qubits.}
\label{fig:chain_length}
\end{figure}

All we now have to do is show how to perform a two-qubit gate, such as a controlled-{\sc not}. The basic way that we plan to do this is show how to entangle the {\sc cu} and the control qubit. The state transfer protocol can then be used to move the {\sc cu} to the target qubit, where it performs a standard one-qubit gate, before moving the {\sc cu} back to the control qubit, and reversing the entangling steps to return the {\sc cu} to its original state (Fig.~\ref{2_qubit_gate}). This is the most involved part of a global control scheme, and requires some modification of the structure already outlined. We now choose to shift the computational qubits such that they are spaced onto every {\em{third}} A qubit, the rest being left in the $\ket{0}$ state\footnote{There is some small reduction in terms of the cost of qubits that can be made here, but it has been neglected for the sake of clarity. This reduction is present in the device structure of Fig.~\ref{fig:chain_length}.}. Having moved the {\sc cu} to the right of the control qubit, the entangling sequence proceeds as follows (reading from left to right):
$$
\alpha^{Y-c}\beta^{c-X}\alpha^{c-H}\beta^{c-Z}\alpha^{c-H}
$$
The second step of this process copies the state of the {\sc cu} onto the (empty) A qubit to its right. This is allowed because it's in a classical state. If the {\sc cu} is absent, the $\beta^{c-Z}$ step does nothing, and the two $\alpha^{c-H}$ terms cancel each other. However, if the {\sc cu} is present, a $Z$ is introduced. Since $HZH=X$, it is possible to see that the {\sc cu} gets deactivated conditionally on the state of the qubit to its left (the control qubit), thus giving the result we require.
\begin{eqnarray}
(\alpha\ket{0_A}+\beta\ket{1_A})\ket{0_B}\ket{0_A}&\rightarrow&(\alpha\ket{0_A}+\beta\ket{1_A})\ket{0_B}\ket{0_A}	\nonumber\\
(\alpha\ket{0_A}+\beta\ket{1_A})\ket{1_B}\ket{0_A}&\rightarrow&(\alpha\ket{0_A}\ket{0_B}+\beta\ket{1_A}\ket{1_B})\ket{1_A}	\nonumber
\end{eqnarray}

There are a number of useful points to notice about this protocol. Firstly, if the qubit that we use as the control is a classical bit (i.e.~a definite $\ket{0}$ or $\ket{1}$), this acts as a flag that indicates whether or not to deactivate the {\sc cu}. Care does need to be taken, however. For example, if we continued with this {\sc cu} and tried to perform a controlled gate with it, then, during the entangling step, the location where the original entangling steps takes place creates a new {\sc cu} as well. Such additional interactions are easily compensated for, provided we remember that they happen. We'll come back to this process of enabling/disabling {\sc cu}s, as it is very useful in error correction procedures. Secondly, if we were to measure, simultaneously, all of the B qubits immediately after the entangling operation, given that we know that every B qubit other than the {\sc cu} is in the $\ket{0}$ state, this acts as a measurement on the {\sc cu} and, therefore, acts as a measurement on the single qubit that was acting as the control qubit. The {\sc cu} can still be recovered afterwards so that we can continue with the computation. While not described in the original proposal, \cite{Benjamin:2002a}, this measurement method is supported by the architecture presented.

This means that we now have a universal set of operations in this globally controlled structure, only requiring initialisation of the {\sc cu}.

\subsection{Minimality of Encoding}

The scheme that we have presented here encodes a single computational qubit in every six physical spins. Slight modification of this idea allows the reduction to ten spins for two computational qubits. It is, naturally, an interesting question as to whether this is the minimal encoding. We can give simple arguments that indicate that this is indeed the case. Let us assume that we wish to keep the idea of a control unit. It is clear that this will have to be able to move relative to the computational qubits. As a result, we must already introduce a doubling of the spins (to divide them into As and Bs).

In addition, we must consider the two-qubit gate. There are two possible concepts as to how this could be implemented. Firstly, we could consider the already outlined mechanism of coherently disabling the {\sc cu}. Since this must be a reversible process, it must leave the information about the original {\sc cu} somewhere so that it can be recovered. Given that we are using qubits\footnote{In \cite{kay-2004-6,kay-2006-73} we reduced the overhead by using higher dimensional systems}, this information must be stored on an additional system, and hence could either be placed on another A qubit, requiring a trebling of the device size (an extra A to store the state of the {\sc cu} and an extra A to give a break between the stored {\sc cu} and the next computational qubit), or by placing it on another B, without increasing the device size. We cannot achieve this for a similar reason to that which is outlined below for the futility of the second mechanism for a two-qubit gate.

The second possibility for implementing a two-qubit gate is to use the {\sc cu} to `pull' a computational qubit along with it, as it moves through the device towards the target qubit. To achieve this, the {\sc cu} must be capable of implementing a set of operations that {\sc swap}s its nearest-neighbours. However, in order to achieve this (without performing the {\sc swap} anywhere else), the two qubits to be {\sc swap}ped must interact. This interaction can only occur through the {\sc cu} because we only have two-body interactions, and hence cannot be controlled by the {\sc cu}. Therefore, this is impossible. 

We are therefore left with the scheme outlined so far. A slight improvement can be made by realising that we only have to perform a single two-qubit gate at any time. As a result, we only need to store an inactive {\sc cu} in one place. Therefore, pairs of qubits can share the region of spins where the {\sc cu} can be stored. This reduces the requirements from 6 physical spins per qubit to 10 spins for every 2 qubits.

\subsection{Device Size}

In order to move to a global control scenario, we have had to switch from a situation where every single qubit would have been a computational qubit, to one where every $6^{th}$ qubit is computational, with an additional knock-on cost to the number of steps required in the protocol. When we want to consider a physical device, enumeration of these costs may be important. In particular, if we know how many qubits we can reasonably build in a system, how many computational qubits can we get out? One hidden cost of the global control scheme is that of edge effects. The state transfer protocol, which smoothly moves the row of A qubits through the row of B qubits assumes an infinitely long chain. If we have a finite size of chain, what happens at the ends is that some states that were stored on A qubits start piling up on qubits labelled by B,
and it is possible that pairs of computational qubits would be adjacent to each other during gate processes, causing additional, unwanted interactions.
To avoid this, it is necessary to ensure that there is sufficient `padding' (i.e.~a large number of qubits in the $\ket{0}$ state at either end of the device) for the A states to move into. This means that to implement $N$ computational qubits, the device needs to contain $12N$ physical spins. 

\section{Error Correction}

If the operations that we perform are perfect, then our scheme is complete -- we can implement a universal set of gates in an efficient manner. In reality we will not be able to implement these operations perfectly and therefore we require error correction \cite{Benjamin:2003b}. At first glance, this is a huge obstacle, for two main reasons.

\begin{figure}[t]
  \begin{center}
    \leavevmode
\includegraphics[width=0.35\textwidth]{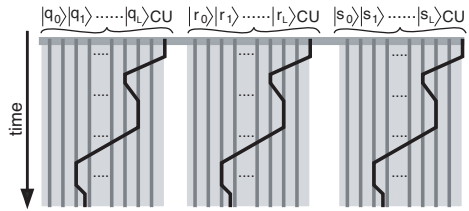}
\end{center}
\caption[Globally controlled structure in error correcting mode]{When in error-correcting mode, a globally controlled architecture must apply the same operations on all blocks of qubits simultaneously. Taken from \cite{Benjamin:2003b}.}
\label{fig:EC_mode}
\end{figure}

Firstly, we have set up our system so that we can implement only one operation at a time. Aharonov and Ben-Or \cite{Aharanov:99} have proved that this is insufficient to be able to implement error correction, and that a degree of parallelism of at least $O(\log(N))$ is required for computation on $N$ qubits. We can consider introducing multiple {\sc cu}s into the device to satisfy this condition, as depicted in Fig.~\ref{fig:EC_mode}. In that case, however, we have lost the ability to implement individual gates within the device, such as during those periods between phases of error correction. To address this problem we will require a method for switching between the two phases where we have different arrangements of {\sc cu}s.

The second problem with error correcting a global control scheme is that traditional descriptions of error correction involve making measurements to determine the locations of errors. These measurement results will be different for each encoded qubit, and hence the required correction will be different as well. As a result, even though we need to run multiple {\sc cu}s in parallel, it appears that they have to do different things! This problem can be circumvented by making the correction procedure coherent.

We need to ensure that we error correct all the qubits in our system. This means not only the computational qubits, but also the {\sc cu}s and the `buffer' qubits -- all those in a classical state that do not play an active role in the computation. To start with, we shall consider that all the classical states are stable, and just describe error correction for the computational qubits.

\subsection{Coherent Correction Procedures}

\begin{figure}[!t]
\begin{center}
\resizebox*{0.5\textwidth}{!}{
\Qcircuit @C=1em @R=.7em {
\lstick{\ket{0}}	& \gate{H}	& \ctrl{9} 	& \qw	 	& \qw	 	& \gate{H} & \ctrl{1}	& \gate{X}	& \ctrl{1}	& \qw		& \ctrl{1}	& \gate{X}	& \ctrl{1}	& \qw		& \ctrl{1}	& \qw		& \ctrl{1}	& \gate{X}	& \ctrl{1}	& \qw	\\
\lstick{\ket{0}}	& \gate{H}	& \qw	 	& \ctrl{8} 	& \qw	 	& \gate{H} & \ctrl{1}	& \qw		& \ctrl{1}	& \gate{X}	& \ctrl{1}	& \qw		& \ctrl{1}	& \qw		& \ctrl{1}	& \gate{X}	& \ctrl{1}	& \qw		& \ctrl{1}	& \qw	\\
\lstick{\ket{0}}	& \gate{H}	& \qw	 	& \qw	 	& \ctrl{7} 	& \gate{H} & \ctrl{7}	& \qw		& \ctrl{3}	& \qw		& \ctrl{1}	& \qw		& \ctrl{5}	& \gate{X}	& \ctrl{4}	& \qw		& \ctrl{6}	& \qw		& \ctrl{2}	& \qw	\\
					& \qw		& \qw		& \qw		& \targ		& \gate{H} & \qw		& \qw		& \qw		& \qw		& \targ		& \qw		& \qw		& \qw		& \qw		& \qw		& \qw		& \qw		& \qw		& \qw	\\
					& \qw		& \qw		& \targ		& \qw		& \gate{H} & \qw		& \qw		& \qw		& \qw		& \qw		& \qw		& \qw		& \qw		& \qw		& \qw		& \qw		& \qw		& \targ		& \qw	\\
					& \qw		& \qw		& \targ		& \targ		& \gate{H} & \qw		& \qw		& \targ		& \qw		& \qw		& \qw		& \qw		& \qw		& \qw		& \qw		& \qw		& \qw		& \qw		& \qw	\\
					& \qw		& \targ		& \qw		& \qw		& \gate{H} & \qw		& \qw		& \qw		& \qw		& \qw		& \qw		& \qw		& \qw		& \targ		& \qw		& \qw		& \qw		& \qw		& \qw	\\
					& \qw		& \targ		& \qw		& \targ		& \gate{H} & \qw		& \qw		& \qw		& \qw		& \qw		& \qw		& \targ		& \qw		& \qw		& \qw		& \qw		& \qw		& \qw		& \qw	\\
					& \qw		& \targ		& \targ		& \qw		& \gate{H} & \qw		& \qw		& \qw		& \qw		& \qw		& \qw		& \qw		& \qw		& \qw		& \qw		& \targ		& \qw		& \qw		& \qw	\\
					& \qw		& \targ		& \targ		& \targ		& \gate{H} & \targ		& \qw		& \qw		& \qw		& \qw		& \qw		& \qw		& \qw		& \qw		& \qw		& \qw		& \qw		& \qw		& \qw	\\
}}
\caption[Coherent feedback in error correction]{Error correction of the Steane [[7,1,3]] code using a coherent feedback process instead of measurement and correction. This circuit must be repeated twice to correct for both $X$ and $Z$ errors.}\label{fig:steane}
\end{center}
\end{figure}
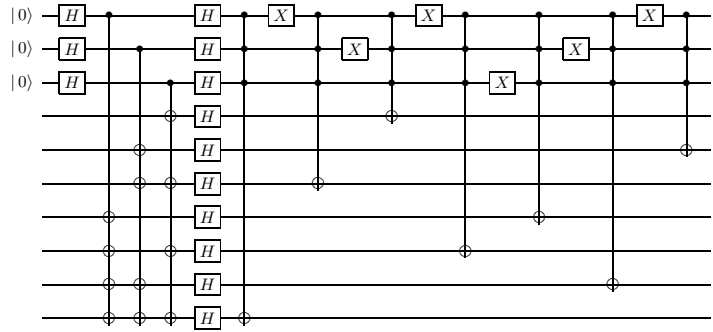

When error correction is performed, the scheme typically follows the process of performing syndrome extraction so that some ancilla qubits contain information on the errors. After that, the ancillas are measured and we act depending on the results in some fixed, logical way. Hence, we might as well include this logic in a quantum circuit that feeds the error information directly back to the encoded qubit. This way, the correction procedure is the same for every encoded qubit, independent of what errors have occurred. The information on what errors occurred is left in the ancilla qubits, which then need to be reset. This idea is illustrated in Fig.~\ref{fig:steane}.

\subsection{Ancilla Reset}

In order to avoid needing to measure and correct, we need to be able to reset qubits that are in an unknown state to the $\ket{0}$ state. To achieve this, we assume that our qubits are not just qubits but have a third level, which we can populate from either of the states $\ket{0}$ and $\ket{1}$. However, we will also assume that this third level, which is at some higher energy than the computational states, has a dissipative decay to one particular state, say $\ket{0}$. This is essentially the same idea as algorithmic cooling \cite{algo_cool}. The reset procedure is denoted in our circuits by $\lightning$.

\subsection{Switchable Parallelism}

\begin{figure}[t]
  \begin{center}
    \leavevmode
\resizebox{0.45\textwidth}{!}{\includegraphics{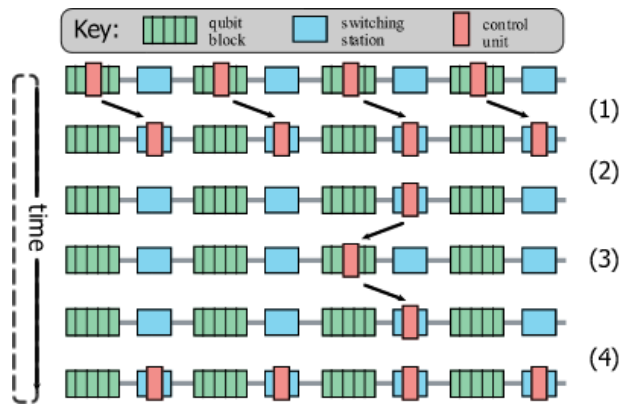}}
\end{center}
\caption[Switchable parallelism]{In order to switch between our algorithm on a quantum computer and error correcting mode, we must switch on and off a regular array of {\sc cu}s, except for one which remains constant. This is achieved by performing the start of a two-qubit gate on a classical state stored at the edge of each encoded qubit. (1) The system starts in a state with one {\sc cu} per switching station, performing operations in parallel on each section of the computer. (2) To switch to an algorithmic step, the {\sc cu}s are moved to the switching stations, and all but one are deactivated. (3) The single {\sc cu} performs an algorithmic step on a set of qubits. (4) The single {\sc cu} returns to its original switching station, and all {\sc cu}s are reactivated for another cycle of error correction. Taken from \cite{Benjamin:2003b}.}
\label{fig:switchable}
\end{figure}

In order to perform computation on our device, we require the ability to switch between two different scenarios. Firstly, we need a {\sc cu} for every encoded qubit, so that we can perform error correction on it. Secondly, we need a single {\sc cu} in the computer so that we can perform the algorithmic part of our computation. Thankfully, we have already seen how we can switch a {\sc cu} off -- we just use the first steps of a two-qubit gate, where the control qubit is a classical state. This classical state indicates whether the {\sc cu} should be left switched on or not. We can therefore consider a slight modification of our device structure. We have many repeating blocks of $L$ qubits, each encoding a single logical qubit. Adjacent to each block is another qubit, which is initialised in a classical state. All of these are set to $\ket{1}$, except for one, which is set to $\ket{0}$. This additional qubit in the block of $L$ is referred to as the Switching Station ({\sc ss}).

Our computer starts off in error-correcting mode, i.e.~with one {\sc cu} for every logical qubit. When we want to perform a computational step, we move the {\sc cu}s to the {\sc ss}s (given the regular spacing of all the {\sc cu}s and {\sc ss}s, this happens simultaneously for all of them) and perform the first step of a two-qubit gate. This deactivates all the {\sc cu}s in a reversible manner in every {\sc ss}, except for the one that is set to $\ket{0}$, leaving a single {\sc cu} for performing the algorithmic steps that we require.

Ideally, between two phases of error correction we would be able to perform an arbitrary computational step. However, given the one-dimensional organisation of our computer, we have to accept that for increasing device size, this is impossible. Instead, when we have to perform gates between distant qubits, we substitute this for a series of {\sc swap} gates to gradually move the qubits closer together, so that they can eventually be interacted. One subtlety, however, is that our single {\sc cu} appears in a single position, and so we need an increasing number of steps just to move it into the correct position. In the same way that we can move a computational qubit along the array, we can move the {\sc cu}'s starting position along the array. We achieve this by understanding how the deactivated {\sc cu}s are stored in a pattern of $\ket{101}$ on physical qubits. Our single-qubit gate protocol allows us to create this pattern in the region where our {\sc cu} remains switched on, and remove it from the next {\sc ss} along.

\subsection{Parallelism for Fault Tolerance}

\begin{figure}[t]
  \begin{center}
    \leavevmode
\resizebox{0.45\textwidth}{!}{\includegraphics{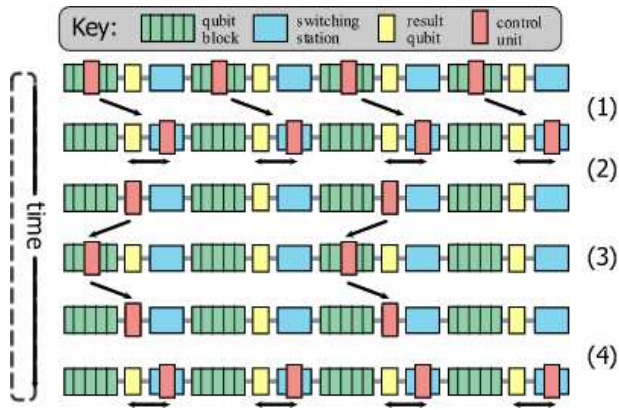}}
\end{center}
\caption[Switching stations]{To generate more complex sets of {\sc cu}s, we perform computation on a switching station (2), which has been classically preprogrammed with sufficient information. The output is placed on a result qubit, which is used to conditionally deactivate sets of {\sc cu}s. At the end of the process, the same result bit is used to reactivate the {\sc cu}s, and the initial computation is inverted (4), or the result bit is reset.}
\label{fig:switchingstations}
\end{figure}

Benjamin originally introduced the idea of {\sc ss}s in \cite{benjamin:2002b}, where he referred to them as sub-computers. Each sub-computer had a unique label, and the {\sc cu}s would perform a computation on each label to decide whether they should be deactivated, allowing arbitrary patterns of {\sc cu}s to be created. The concept of {\sc ss}s is much more limited -- we only need two different patterns of {\sc cu}s. As a result, these {\sc ss}s only require a fixed proportion of the device size, whereas the labels for the sub-computers grew with $\log(N)$. While this is still a relatively modest cost, the computation time to create arbitrary patterns of {\sc cu}s must require $O(N)$ steps.

The next logical question is whether there are fixed patterns of {\sc cu}s that we can usefully create, while still requiring only a fixed proportion of the device size, and only requiring a fixed amount of the computation time. In particular, we would like to consider the configurations for fault-tolerance (FT). In this scenario, we desire the ability to switch from a single {\sc cu} to sets of {\sc cu}s that are active every $L^n$ encoded qubits, where $n$ is an integer from 0 (the original {\sc ec} pattern) up to some maximum level, $p-1$. This is because an encoded qubit consists of $L$ computational qubits. Hence, when we concatenate a single level of code, the encoded qubits that we now produce consist of $L$ of the originally encoded qubits. This continues up the levels of concatenation, yielding the indicated power law. The number of levels of concatenation, $p$, is independent of the number of qubits that we want to perform computation on, it only depends of the accuracy to which we desire our final computation. Hence, within our {\sc ss} we could envisage $p$ qubits, each indicating whether a particular {\sc cu} should be turned on at each level of concatenation. This would require a minimum of computation at each step (one gate), and still only requires a fixed proportion of the device size.

This expanded {\sc ss} is actually more powerful than we require, and could be useful for some more advanced ideas, such as super-{\sc cu}s. The rationale behind a super-{\sc cu} is that when we operate on concatenated codes, most of the operations that we perform can be done so bitwise. Instead of using a single {\sc cu} and repeating the same operation many times, it would be more sensible to have a super-{\sc cu}, which consists of a dense block of {\sc cu}s, allowing us to perform these bitwise steps with a single command. However, for the most simple form of fault-tolerance, this expanded {\sc ss} is more complex than we require. Instead of $p$ qubits, we can reduce it to $\lceil\log_2(p+1)\rceil$ qubits, and a small additional computation. The point here is that we can organise it so that when we require a {\sc cu} every $L^n$ qubits instead of every $L^{n-1}$ qubits, we only have to switch off {\sc cu}s (there is no need to reactivate any). As a result, we just need the label of the {\sc ss} to say at which level the corresponding {\sc cu} gets switched off. Hence, we only have to encode the numbers $0$ to $p-1$, which can be achieved in $\lceil\log_2(p)\rceil$ qubits. Of course, we also have to store the single extra number ($p+1$) that marks the location of the single {\sc cu} which we retain for performing the algorithmic steps of the computation.


\begin{table}
\begin{tabular}{|c|r|r|}
\hline
& $b_n=0$ & $b_n=1$ \\
\hline
Initial Step ($n=1$) &
\hspace{1.5cm}
\Qcircuit @C=1em @!R{
\lstick{\ket{a_1}} & \ctrl{1} & \ctrl{2} & \qw & \qw \\
\lstick{\ket{r}=\ket{0}} & \targ & \qw & \qw & \qw \\
\lstick{\ket{c_1}=\ket{0}} & \qw & \targ & \gate{X} & \qw\\
} &
\hspace{1.5cm}
\Qcircuit @C=1em @!R{
\lstick{\ket{a_1}} & \ctrl{2} & \qw \\
\lstick{\ket{r}=\ket{0}} & \qw & \qw \\
\lstick{\ket{c_1}=\ket{0}} & \targ & \qw
}	\\
& &\\
\hline
Intermediate Steps &
\Qcircuit @C=1em @!R{
\lstick{\ket{c_{n-1}}} & \ctrl{1} & \ctrl{1} & \ctrl{3} & \qw \\
\lstick{\ket{a_n}} & \ctrl{1} & \ctrl{2} & \qw & \qw \\
\lstick{\ket{r}} & \targ & \qw & \qw & \qw \\
\lstick{\ket{c_n}=\ket{0}} & \qw & \targ & \targ & \qw
} &
\Qcircuit @C=1em @!R{
\lstick{\ket{c_{n-1}}} & \ctrl{1} & \qw \\
\lstick{\ket{a_n}} & \ctrl{2} & \qw \\
\lstick{\ket{r}} & \qw & \qw \\
\lstick{\ket{c_n}=\ket{0}} & \targ & \qw
} \\
& &\\

\hline
Final Step ($n=\log_2(p)$) &
\Qcircuit @C=1em @!R{
\lstick{\ket{c_n}} & \ctrl{2} & \qw \\
\lstick{\ket{a_n}} & \qw & \qw \\
\lstick{\ket{r}} & \targ & \qw
} &
\Qcircuit @C=1em @!R{
\lstick{\ket{c_n}} & \ctrl{1} & \qw \\
\lstick{\ket{a_n}} & \ctrl{1} & \qw \\
\lstick{\ket{r}} & \targ & \qw
} \\
& &\\
\hline
\end{tabular}\caption[Algorithm to test if $a\geq b$]{Quantum algorithm, using ancillas $c_i$ to determine if $a\geq b$. $a_x$ is the $x^{th}$ most significant bit of the number $a$, which is stored in the {\sc ss}. $b$ represents the level of concatenation for which we wish to activate the {\sc cu}s.}\label{fig:FT_algo}
\end{table}

We must still show how the actual process of activation/deactivation of {\sc cu}s occurs. To achieve this, we will start with a {\sc cu} active in every {\sc ss}\footnote{If we are just switching from one level of concatenation to another, it will generally more efficient to by-pass this step, and just use the already active {\sc cu}s.}, and perform a small (reversible) computation on the label of the {\sc ss}. This computation will output a single bit onto an ancilla qubit, which can be used for the activation/deactivation procedure as before. We give the required computation in Tab.~\ref{fig:FT_algo}. The concept of the circuit is quite simple to understand. We wish to determine if $b>a$. Starting with the most significant bit of each, if $b_1>a_1$, then $b>a$ and there is no need to continue Similarly, if $b_1<a_1$, there is no need to continue. We then move to the next most significant bit. We only compare bit $x$ if all the more significant bits of $a$ and $b$ are equal. This continues until termination of the sequence, either because the answer is determined, or because we have run out of bits (in which case we must also know the answer, but the terminating step is consequently slightly different).

We therefore see that it is possible to generate sufficient parallelism for fault-tolerance. In subsequent sections, we will ensure that we have taken into account all necessary considerations by deriving a fault-tolerant threshold for our scheme. However, before we do this, we must consider all the other qubits in the system, not just the computational qubits.

\section{Error Correction of the Control Units} \label{sec:ec_CU}

We have now presented a scheme that appears to be able to perform arbitrarily accurate computation on the one-dimensional globally controlled array, even in the presence of (small) errors. However, this scheme has implicitly assumed the stability of all the classical states in the system - all the padding $\ket{0}$ states, the labels in the {\sc ss}s and the {\sc cu}s. If any of these become corrupted, the whole computation can become corrupted. As a first step towards protecting these, we can simply state that our regular device structure guarantees the locations of these classical states in a periodic way. Hence, potentially, we could create global pulses that just address these. For example, the B qubits should always be classical states (unless we're in the middle of performing a two-qubit gate). By applying regular measurement to these classical states, we get a Zeno effect which prohibits transition of a classical state into its complement. In the long-run, this is not sufficient for arbitrarily accurate computation, but may delay the requirement for more aggressive schemes of error correction.

Eventually, we require the ability to perform error correction on the {\sc cu}s. The first step in doing this is to acknowledge that these are classical states and, hence, we only have to protect them against bit-flips, and not phase-flips. As such, we can use a classical repetition code to protect the information. We still intend to perform our computation with single {\sc cu}s, but at the error correcting stage we will switch on different sets of triples of {\sc cu}s. These triples will compare themselves to each other and attempt correction. This step is somewhat non-trivial because we must allow the potentially faulty {\sc cu}s to control their own actions, and yet we still need to be sure that the correction will result. One way in which this can be achieved was first presented in \cite{Kay:2005c}.

So, instead of using a single {\sc cu} for a computation, we shall now use three of them. We do not intend to perform any part of the algorithm with all three {\sc cu}s present, as this will be less efficient than switching off two of them. We can initially align these with computational qubits. The patterning needs to be chosen with care in order to minimise the complexity of the protocol. We target an operation on a single qubit by applying a controlled-phase gate ($CP$) with each of the three {\sc cu}s, separated by the single qubit rotations $U_1$, $U_2$, $U_3$ and $U_4$ applied to all the $A$ qubits. The resulting evolution on the targeted qubit is
$$
U_1ZU_2ZU_3ZU_4.
$$
Any qubits that are far enough away from the {\sc cu}s will not be affected by the $CP$s, and hence will be subject to the evolution $U_1U_2U_3U_4$, which we select to be the identity transformation i.e.~$U_4=U_3^\dagger U_2^\dagger U_1^\dagger$. We need to create sequences such that we can apply an arbitrary rotation to the qubit we want, but do nothing to any of the other qubits. This necessarily includes qubits that get affected by one or two of the $CP$s. The first step is to select a patterning of the {\sc cu}s such that no qubit ever experiences 2 $CP$s (unless an error has occurred). The simplest example is to align the {\sc cu}s with qubits $q_1$, $q_2$ and $q_4$ (note the gap, $q_3$). 

In the fault-tolerant scenario that we have been describing, the most efficient way of introducing these three {\sc cu}s is by relabelling the switching stations using the following protocol. If the label was non-zero, add 1 to the value (these correspond to the qubits $q_1$ for each block). Label the zero-valued {\sc ss}s that are 1 and 3 {\sc ss}s away from these relabelled {\sc ss}s with the number 1 (corresponding to qubits $q_2$ and $q_4$). So, by deactivating all {\sc cu}s in regions controlled by {\sc ss}s with 0 labels, we get the regular patterning that is required, and we can still access the parallelism required for fault-tolerance. In fact, by repeating this procedure at every level of concatenation, this enables patterns of {\sc cu}s for fault-tolerance at a cost of a single extra bit in each {\sc ss}.

The 1-2-4 arrangement of {\sc cu}s can still cause unavoidable applications of single $CP$ gates,
\begin{eqnarray}
&U_1ZU_1^\dagger&	\nonumber\\
&U_1U_2ZU_2^\dagger U_1^\dagger&	\nonumber\\
&U_1U_2U_3ZU_3^\dagger U_2^\dagger U_1^\dagger.&	\nonumber
\end{eqnarray}
Clearly, if we apply the sequence {\em{twice}}, all of these will be removed, while the targeted qubit experiences the evolution
$$
U_1ZU_2ZU_3ZU_3^\dagger U_2^\dagger ZU_2ZU_3ZU_3^\dagger U_2^\dagger U_1^\dagger
$$
which, given that we have a free choice of $U_1$, $U_2$ and $U_3$, must contain sufficient freedom to create any single-qubit rotation that we desire (up to a global phase).

Now that we are using a {\sc cu} that has some redundancy in it, there is enough information to be able to correct for errors. There are two stages involved in making use of this information. Firstly, we must perform a syndrome extraction, placing information about any errors that have occurred on an ancilla qubit (which would otherwise have been a computational qubit). Secondly, we must feed back from this ancilla to be able to correct the faulty part of the {\sc cu}.

If one of the {\sc cu}s suffers a bit flip, then only the targeted qubit will be affected. We can therefore neglect all other qubits, and just concentrate on the ancilla that we will be targeting, and which will initially be in the state $\ket{0}$. For the three bits that can get flipped, the resulting evolution will be one of
\begin{eqnarray}
&U_1U_2ZU_3ZU_3^\dagger ZU_3ZU_3^\dagger U_2^\dagger U_1^\dagger,&	\nonumber\\
&U_1ZU_2U_3ZU_3^\dagger U_2^\dagger ZU_2U_3ZU_3^\dagger U_2^\dagger U_1^\dagger,&	\nonumber\\
\text{or}&U_1ZU_2ZU_2^\dagger ZU_2ZU_2^\dagger U_1^\dagger. &	\nonumber
\end{eqnarray}
Using either $U_2=\identity$ or $U_3=\identity$, the evolution when the {\sc cu} has no error is $\identity$. The results if there have been errors are shown in Table \ref{tab:syndrome}, where
$$
V_n=U_1ZU_nZU_n^\dagger ZU_nZU_n^\dagger U_1^\dagger.
$$
$n$ is either 2 or 3, where $U_n\neq\identity$. We thus have free choice of $U_1$ and $U_n$ to make this remaining evolution $X$ (for example, $U_1=\identity$ and $U_n=e^{-iX\pi/4}$), which enables the error syndrome to be placed on the ancilla. Similarly, we can create the evolution $Z$ just by changing $U_1$ to $H$. If we wanted to create the Hadamard gate, we would set
$$
U_1=\frac{1}{\sqrt[4]{8}}\left(\begin{array}{cc}
\sqrt{\sqrt{2}+1} & -\sqrt{\sqrt{2}-1} \\
\sqrt{\sqrt{2}-1} & \sqrt{\sqrt{2}+1}
\end{array}\right).H
$$

\begin{table}[!t]
\begin{center}
\begin{tabular}{|c|c|c|}
\hline
Error occurs on {\sc cu}:	& $U_2=\identity$	&	$U_3=\identity$	\\
\hline
none	&	$\identity$	& $\identity$	\\
1	&	$V_3$	&	$\identity$	\\
2	&	$V_3$	&	$V_2$	\\
3	&	$\identity$	&	$V_2$	\\
\hline
\end{tabular}
\caption[Error detection of {\sc cu}s]{The evolution that can occur if an error affects a single {\sc cu} out of the three. By selecting the single-qubit rotations suitably, syndrome extraction can be performed.}
\label{tab:syndrome}
\end{center}
\end{table}

\begin{figure}[!t]
\begin{center}
\includegraphics[width=0.45\textwidth]{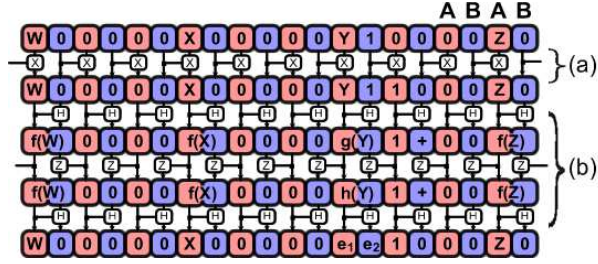}
\caption[Error correction of {\sc cu}s]{Method for feeding back error syndrome to {\sc cu}s. The specification differs from the first half of Fig.~\ref{2_qubit_gate} because the steps in part (a) are now controlled by the 2 {\sc cu}s that are not being corrected. The first step has also been removed. Part (b) is applied to the {\sc cu} that is being corrected only.}
\label{2_qubit_gateb}
\end{center}
\end{figure}

How can this information be used to correct the right error? If we apply $V_3=X$, i.e.~we have set $U_2=\identity$, then the target (ancilla) qubit is flipped if there is an error on either of the first two {\sc cu}s. We then use the circuit in Fig.~\ref{2_qubit_gateb} to feed back the error syndrome from the ancilla to the first {\sc cu}, where part (a) is controlled by both {\sc cu}s 2 and 3. If the error occurred on the second {\sc cu}, then the feedback process won't occur. Therefore, we can correct for a single error on any of the {\sc cu}s by repeating the process.

Alternatively, we could target the same ancilla with pulse sequences such as $(V_2=H).(V_3=Z).(V_2=H)$, which would flip the ancilla only if an error has occurred on the second {\sc cu}. In the previous method, we created the local pattern of 2 $\ket{1}$s on neighbouring A qubits which is required to feed information back to the {\sc cu}s by using the {\sc cu}s. This is susceptible to error if a nearby B qubit that should be a $\ket{0}$ gets flipped to a $\ket{1}$ (the local pattern can be created in several places). However, this alternative technique allows us to make one of those $\ket{1}$s a fixed classical state (within the {\sc ss}, say), and then we need concentrate on only flipping the one ancilla next to it. In this case, if only a single error occurs somewhere, correction is localised in the right area of the device, and is stable against other B qubits being flipped. 

\subsection{Stability of the Classical States}

Now that we know how to keep the {\sc cu}s stable, it is relatively easy to see how to keep the other classical states stable. We don't even need error correction, we just need to reset the qubits conditional on the presence of the now perfect {\sc cu}. If the reset procedure is imperfect, then it just gets corrected in the next round of resets. This involves two steps, one which corrects those on $A$ qubits and another that corrects those on $B$ qubits. Both of these will occur in the regime where there is one {\sc cu} activated for every {\sc ss}. Resetting the $A$ buffers is simple - we just move the {\sc cu} next to each of them and send out the reset command, much as we would for a single-qubit gate. This requires a fixed number of steps, independent of device size.

To correct the $B$ buffers, we move the {\sc cu} adjacent to the result qubit in the {\sc ss} (the one which is used as the control bit), and switch it to the $\ket{1}$ state. We also do that to an adjacent $A$ qubit (i.e.~the nearest physical $A$ qubit, not the next computational qubit). This creates the unique local patterning which is used to feed operations back onto the $B$ qubits (see Fig.~\ref{2_qubit_gate}). Hence we move all the buffer qubits between these two $\ket{1}$s and reset them. One potential risk is an interaction with the {\sc cu} when it is adjacent to one of the $\ket{1}$s during the feedback process. Since the mechanism for the reset procedure will depend on the physical implementation, this question is difficult to answer, but we can at least say that if we were applying a unitary operation, there is no additional interaction.

One might think that the unique patterning, the same as found in the two-qubit gate, could be used to deactivate the {\sc cu}, meaning that all the $B$ qubits should be in the $\ket{0}$ state, so we could globally apply the reset procedure to all of them. Unfortunately, this process would also move the imperfections of the $B$ qubits onto the $A$ qubits, and, in turn, move some of the computational qubits onto $B$ qubits (if there are errors on some of the $A$ buffers). This reminds us that if there are imperfections in the $B$ qubits, these will be mapped onto the $A$ qubits, including the computational qubits. However, error correction will correct for these provided they occur sufficiently infrequently.

Having reset the buffer qubits, we need to reset, or otherwise stabilise, the qubits in the {\sc ss}s. We can use the different levels of concatenation in the fault-tolerant scenario to our advantage for performing resets of different organisations of qubits. Consider the scenario where we have a {\sc cu} enabled for every $L$ {\sc ss}s. Between their own {\sc ss}s, there are $L-1$ {\sc ss}s which all store the value 0 and have deactivated {\sc cu}s. Thus, we can reset all of these. Note, however, that they can't reset their own states. At higher levels of concatenation, there is a {\sc cu} for every $L^i$ {\sc ss}s. Regularly spaced between each of these are $L-1$ {\sc ss}s which contain the number $i-1$ and deactivated {\sc cu}s that have not been corrected yet. So, we can move the active {\sc cu}s along and reset all of these. Eventually, at the top level of concatenation, there are some switching stations that have never been reset. We cannot correct them with the single {\sc cu} (errors would build up too quickly as the device size scales), so this would appear to be a problem. However, it is not. We have gone to a level of concatenation that is as good as we need i.e.~it is sufficiently good that it is safe to assume that the top level of concatenation is stable. Hence, it is acceptable for these {\sc cu}s to correct their own {\sc ss}s -- the effect that we are concerned with is precisely the same effect as the termination of the hierarchy of concatenation instead of continuing it {\em ad infinitum}.

\section{Error Correction of Quantum Cellular Automata}

As we have presented it here, our physical model corresponds in a natural way to a particular realisation of quantum computation -- optical lattices. In particular, we have assumed we can perform operations on the A qubits, controlled by the B qubits to their right (for example). Another global control scenario which, again, has a physical counterpart, is that our qubits can only determine the total spin of their neighbours, $\pm 1, 0$ \cite{Benjamin:2001b,Benjamin:2002a,BenjaminBose1}. Our global pulses then take the form of $A^X_0$, which causes all the $A$ qubits to flip their value if the total spin of their neighbours is 0 (i.e.~one $B$ qubit is spin up, and the other is spin down). This is precisely the model of quantum cellular automata (see \cite{QCA} and references therein). All of the above work on error correction, {\sc cu}s, {\sc ss}s etc.~can be developed in this scenario. However, there is one vital difference, in that all schemes that have so far been discovered require an encoding of the computational qubits across several physical qubits. This introduces an additional complication, which was carefully circumvented by our choice of physical model -- the computational basis itself must be stabilised. In particular, the logical qubits exist inside a subspace of these physical qubits, and if a faulty pulse sequence causes a departure from this subspace, an error correcting code can do nothing about this. Some measures can be taken, such as stabilisation of the basis via the Zeno effect. This can potentially work to a similar level as error correction, in that we may be able to correct for a single error occurring between two Zeno pulses. However, it is still possible that two errors could occur, and ruin the computation. Hence, these ideas are not sufficient for a fully fault-tolerant scenario. Even after the work of the next section, therefore, the possibility of fault-tolerance in a quantum cellular automata remains an interesting open question. It may be solvable using the techniques of \cite{Bacon:99, Bacon:00, Lidar:02, Lidar:04}, where it is shown that it is possible to keep encoded qubits within a particular subspace, and especially \cite{mohseni:05} where this is done under a form of global control. We have not, as yet, explored this possibility.

We can even justify that, using the concept of a {\sc cu}, we {\em{must}} use an encoded basis for the CA model. This argument is simply a symmetry argument. Let us assume that we can encode the qubits and the {\sc cu} in single spins. Since all the $A$ qubits will behave the same, and the {\sc cu} must do something different to the computational qubits, we can envisage placing the computational qubits on $A$s, and the {\sc cu} on a $B$. We then need to move the {\sc cu} relative to the computational qubits.
$$
\begin{array}{cccccccccc}
A & B & A & B & A & B & A & B & A & B	\\
0 & 0 & 0 & 0 & 0 & 1 & 0 & 0 & 0 & 0
\end{array}
$$
If we were to send a pulse $B_x^y$, then clearly we cannot do anything on just the {\sc cu}. If we send a pulse $A_x^y$, then we can perform operations on the neighbours of the {\sc cu} by setting $x=0$. However, by symmetry, both neighbouring $A$ qubits perform the same operations i.e.~if we can find a sequence of pulses to propagate the {\sc cu} to the left, it also moves to the right. We can either choose to use this symmetry in our system, which requires us to double its size, so the {\sc cu} is encoded on 2 spins, or we can break the symmetry with respect to $A$ and $B$ by choosing an encoding of the computational basis over an even number of spins. Raussendorf and others have showed another way to use the symmetry of the system without requiring a {\sc cu} in the system (effectively by using edge effects to replace the {\sc cu}) \cite{raussendorf:05,fitz:06}.

\section{Deriving Fault-Tolerance and Constructing the Circuits}\label{chap:6}

Having presented a globally controlled architecture that supports fault-tolerance, we would now like to calculate the error rate below which the error improves with each round of concatenation, and hence below which we can make our computation arbitrarily accurate. The approach that we take is to follow the proofs of \cite{gottesman:2005}, which provide a rigorous derivation of a threshold. The authors then proceed to evaluate this threshold under the assumptions of arbitrary parallelism, non-nearest neighbour interactions and the ability to perform measurements. Under the assumption of global control, we have to remove all these simplifications, so we expect the threshold to be significantly worse. As with initial threshold estimates, however, the important point is not how large or small such an error rate is, but that a critical error rate does exist. The ability to derive a threshold in this model is also very useful because it is closely related to quantum cellular automata \cite{QCA}, and whether fault-tolerant computation can be achieved with them. However, as previously described, there are still outstanding issues related to the fault-tolerance of a quantum cellular automaton.

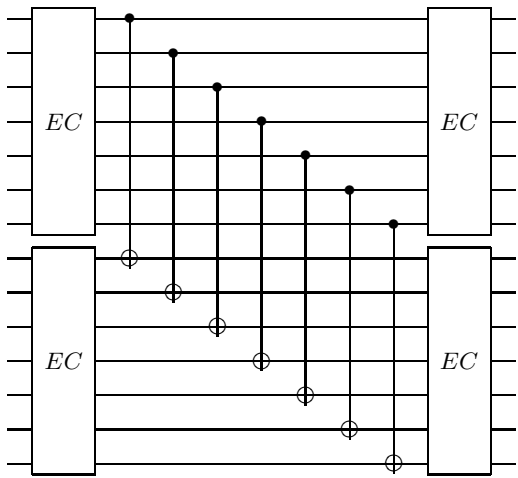
\begin{figure}
\begin{center}
\centerline{
\Qcircuit @C=1em @R=.5em {
& \multigate{6}{EC} & \ctrl{7} & \qw & \qw & \qw & \qw & \qw & \qw & \multigate{6}{EC} & \qw \\
& \ghost{EC} & \qw & \ctrl{7} & \qw & \qw & \qw & \qw & \qw & \ghost{EC} & \qw \\
& \ghost{EC} & \qw & \qw & \ctrl{7} & \qw & \qw & \qw & \qw & \ghost{EC} & \qw \\
& \ghost{EC} & \qw & \qw & \qw & \ctrl{7} & \qw & \qw & \qw & \ghost{EC} & \qw \\
& \ghost{EC} & \qw & \qw & \qw & \qw & \ctrl{7} & \qw & \qw & \ghost{EC} & \qw \\
& \ghost{EC} & \qw & \qw & \qw & \qw & \qw & \ctrl{7} & \qw & \ghost{EC} & \qw \\
& \ghost{EC} & \qw & \qw & \qw & \qw & \qw & \qw & \ctrl{7} & \ghost{EC} & \qw \\
& \multigate{6}{EC} & \targ & \qw & \qw & \qw & \qw & \qw & \qw & \multigate{6}{EC} & \qw \\
& \ghost{EC} & \qw & \targ & \qw & \qw & \qw & \qw & \qw & \ghost{EC} & \qw \\
& \ghost{EC} & \qw & \qw & \targ & \qw & \qw & \qw & \qw & \ghost{EC} & \qw \\
& \ghost{EC} & \qw & \qw & \qw & \targ & \qw & \qw & \qw & \ghost{EC} & \qw \\
& \ghost{EC} & \qw & \qw & \qw & \qw & \targ & \qw & \qw & \ghost{EC} & \qw \\
& \ghost{EC} & \qw & \qw & \qw & \qw & \qw & \targ & \qw & \ghost{EC} & \qw \\
& \ghost{EC} & \qw & \qw & \qw & \qw & \qw & \qw & \targ & \ghost{EC} & \qw
}}
\caption[Schematic circuit construction for fault-tolerance]{Schematic construction for controlled-{\sc not} gate on Steane [[7,1,3]] code, with inputs and outputs surrounded by error correcting circuits.}\label{fig:EC_concept}
\end{center}
\end{figure}

The calculation of a fault-tolerant threshold proceeds approximately as follows. First of all, we select a universal (generally over-complete) set of gates that we want to work with. Then we construct our error correcting circuit out of these gates, ensuring that (in this case) a single error occurring within the circuit propagates to no more than one output qubit. We then take each gate in our universal set (constructed to act on an encoded qubit), and apply the error correction circuit to all the logical input and output qubits. This idea is shown schematically in Fig.~\ref{fig:EC_concept} for the controlled-{\sc not} gate acting on the Steane [[7,1,3]] code, where the controlled-{\sc not} is implemented by applying it bitwise on the physical qubits.

The construction of our circuits has to take into account all the mechanisms that we require, such as the {\sc cu}s, the buffer qubits, nearest-neighbour interactions etc. One of the primary simplifying assumptions that we will make is that the {\sc cu}s and buffer qubits are stable. We can do this because we will be able to construct a fault-tolerant scheme for these, with their own threshold $\epsilon_c$. We expect this threshold will be much larger than the threshold for the computation that we wish to implement (primarily because all the states are classical, so we just have to use a repetition code, which is vastly simpler). Since this threshold is much larger, we can assume that the time required to implement it is a negligible fraction of the computation time, and hence we can just neglect it.

In order to take into account the physical structure of our device, we shall formulate all the gates in terms of nearest-neighbour interactions, with the additional caveat that gate $n+1$ must be applied to a qubit that either is, or is adjacent to, the qubit that gate $n$ was applied to. In this way, we don't have to worry about the {\sc cu} running up and down the whole time, it just moves to its neighbour, costing a maximum of 3 {\sc swap} operations. As such, the {\sc swap} gates that we will add to the circuits are only present to correctly evaluate the number of `wait' operations -- we do not actually implement the {\sc swap} operations. As such, the propagation of errors through these gates is irrelevant. In fact, this is very important because otherwise these {\sc swap} gates could cause the errors to propagate to every qubit involved, which is certainly undesirable.

We also have to remember that we can only apply one gate at a time in each error correcting block, and that we have to apply the `do nothing' operation on the rest of the qubits. This adds significantly to the number of steps in the scheme.

Now that we have written out the circuits under these constraints, we count up the number of locations in which an error can occur in each of the circuits. We then take just the gate (plus the error corrections) that has the largest location count ($=A$), as this will be the one the determines the threshold. Typically, this gate will be the one with the most inputs and outputs. If errors occur with probability $\epsilon$ independently at each location, then the probability of $n$ errors occurring within the circuit is
$$
\epsilon^n\binom{A}{n}.
$$
The motivation for the construction demonstrated in Fig.~\ref{fig:EC_concept} now becomes clear -- it was proven in \cite{gottesman:2005} that all single errors get corrected by these circuits. The probability of an error on the logical qubits is thus the probability that the error does not get corrected,
$$
\epsilon^{(1)}\leq\sum_{n=2}^A\epsilon^n\binom{A}{n}\approx \epsilon^2\left(\binom{A}{2}+\epsilon\binom{A}{3}\right).
$$
The approximation is valid because we expect (as will be confirmed in Sec.~\ref{sec:classical}) that $\epsilon$ is a small quantity. The computation can be performed to arbitrary accuracy provided $\epsilon^{(1)}<\epsilon$, so equality gives an error correcting threshold, $\epsilon_0$. The threshold can then be improved by enumerating the number of benign locations. These are the pairs of locations at which errors can occur and the logical qubits are still corrected. The first term in the expansion is therefore reduce from $\epsilon^2\binom{A}{2}$ to $\epsilon^2B$.
\begin{eqnarray}
\epsilon_0&=&\epsilon_0^2\left(B+\binom{A}{3}\epsilon_0\right)	\nonumber\\
&=&\frac{\sqrt{B^2+4\binom{A}{3}}-B}{2\binom{A}{3}}	\label{eqn:benign_thresh}
\end{eqnarray}

We intend to calculate the threshold for one particular error correcting code, the Steane 7-qubit code. Primarily, this is because the majority of gates can be performed on encoded qubits in a bit-wise manner, making them very simple. Secondly, this will give us a useful point of comparison with \cite{gottesman:2005}, which also uses this code.

The aim of the remainder of this section is to describe general tactics that allow us to take a specific gate which acts on an encoded qubit and show how it can be performed, ensuring that if only a single error occurs in the circuit, only a single physical qubit from each encoded qubit is affected on the output (of course, we don't mind what happens to ancillas, provided these errors do not propagate to the encoded qubit).

Naturally, some of our gate constructions are automatically fault-tolerant. In particular, any bitwise gates are fault-tolerant because there are no operations that can transfer an error from one physical qubit to another in the same encoded qubit. This highlights the reason for choosing the Steane code -- a universal set of gates exists where only a single gate cannot be applied in a bitwise manner. Note that the gate $N$, which we introduce in the next section allows other gates to be applied in a bitwise manner.

\subsection{Propagation of Errors}

In order to ensure that our gate constructions are fault-tolerant, it is important to understand how errors propagate between qubits. The following identities may be useful:
\begin{eqnarray}
\begin{array}{c}
\Qcircuit{
& \gate{X} & \ctrl{1} & \qw \\
& \qw      & \targ  & \qw   \\
}\end{array} &=&
\begin{array}{c}
\Qcircuit{
& \ctrl{1}	& \gate{X} & \qw \\
& \targ		& \gate{X} & \qw \\
}
\end{array} \label{eqn:bitpropagate}\\
\begin{array}{c}
\Qcircuit{
& \gate{H} & \ctrl{1}	& \gate{H} & \qw \\
& \gate{Z} & \targ  	& \qw & \qw  \\
}\end{array} &=&
\begin{array}{c}
\Qcircuit{
& \qw & \qw & \targ & \qw & \qw \\
& \gate{Z} & \gate{H} & \ctrl{-1} & \gate{H} & \qw
}\end{array}	\label{eqn:phasepropagate}\\
&=&\begin{array}{c}
\Qcircuit{
& \gate{H} & \ctrl{1}	& \gate{H} & \gate{X} & \qw \\
& \qw & \targ  	& \qw & \gate{Z} & \qw \\
}\end{array}	\nonumber
\end{eqnarray}
These essentially state that bit-flip errors propagate from control to target, whereas phase-flip errors propagate in the opposite direction. Hence, if we're going to use some ancillas in the gate constructions, we will only have to ensure they're correct with respect to one type of error, a more realistic task than protecting them against all errors.

\subsection{Cat States and Majority Voting} \label{sec:cat}

The significant problem for fault-tolerant circuits arises when we have to output multiple operations onto a single ancilla, which is then used to feedback information onto the original qubits. See, for example, Fig.~\ref{fig:steane}. In this example, syndrome extraction is achieved by performing four controlled-{\sc not} gates, each controlled by the same ancilla (initially in the $\ket{0}+\ket{1}$ state), targeting the four different qubits from which we are extracting the syndrome. If an error occurs on this ancilla after the first controlled-{\sc not}, for example, then it can feed through the other gates to affect the other three qubits.

This problem is circumvented by replacing the single ancilla with four ancillas, prepared in a `cat' state, $\ket{0000}+\ket{1111}$ (named for Schr\"odinger's cat). Each of the four controlled-{\sc not}s is then controlled by a different ancilla. From there, we can take a vote between the four ancillas, onto a fifth, as to what the correct value is, and this can then be used as the control for the feedback operation. There are two different types of vote that we use, depending on the situation. We refer to these as weak and strong majority voting.

\subsubsection{Weak Majority Voting}

In the example discussed so far, we wanted to calculate the single bit which was the syndrome of a set of operators. We divided this into four separate operators, and the result that we required was the binary sum of these operators, since we only have to take into account the possibility that there is a single error anywhere. In fact, this is most easily achieved using a slightly modified version of the cat state,
\begin{equation}
\frac{1}{\sqrt{2}}H^{\otimes 4}(\ket{0000}+\ket{1111}).
\label{eqn:evenbits}
\end{equation}
This state is an equal superposition of all 4-bit strings with an even weight (i.e. an even number of 1s). If the four target qubits are in the correct $z$-state, then this state is invariant. However, if one of them is faulty, one of the ancilla qubits is flipped and we have an equal superposition of all 4-bit strings with odd weight. Therefore, this difference is very easy to detect, each ancilla performs a controlled-{\sc not} onto a fifth ancilla. If a single error occurs at this stage, then it is only this fifth ancilla which matters.

We now just have to be sure that the state in Eqn.~(\ref{eqn:evenbits}) can be created fault-tolerantly from $\ket{0000}$. Note that phase errors in the final state do not matter to us. We start by creating a cat state by applying $H$ to a single qubit, and then performing controlled-{\sc not}s. If a bit-flip has occurred on the control qubit, then this propagates to all the target qubits. However, we will then apply a Hadamard to all the qubits, and hence convert these into phase-flip errors. These do not propagate to the computational qubits, only affecting the single ancilla bit at the end. If a phase-flip has occurred on one of the targets before the controlled-{\sc not}, then the ancilla was in the $\ket{0}$ state, and hence phase errors are irrelevant. Hence, if a single error occurs, it results in bit-flips on no more than one qubit in the state of Eqn.~(\ref{eqn:evenbits}).

This operation is particularly apparent in the circuit for error correction, Fig.~\ref{fig:FT_EC}.

\subsubsection{Strong Majority Voting}

The other situation that we are interested in, although not explicitly given in Fig.~\ref{fig:FT_EC}, we refer to as strong majority voting. In this situation, we wish to detect the error on a particular qubit several times. For example, in Fig.~\ref{fig:steane}, a particular computational qubit is targeted by three controlled-{\sc not}s performing the syndrome extraction. If it were to suffer an error between these three, the error could propagate to a second qubit due to an incorrect determination of the syndrome. In this case, we want to provide a constant value for the syndrome extraction steps, stored on an ancilla. If the ancilla gets flipped at some stage, then the encoded qubit would be falsely corrected, but it would only introduce a single error, not two.

This proceeds by forming a cat state on several (say 4) ancillas, $\ket{0000}+\ket{1111}$. Controlled-{\sc not} gates are then applied, each controlled by a different ancilla and targeting the same qubit. These qubits can then be used to vote, onto a fourth ancilla, as to whether the qubit had suffered an error. If the state is more than a single flip away from the cat state, then it was caused by a fault on the computational qubit. An alternative methodology is depicted in Fig.~\ref{fig:strong_vote}.

\begin{figure}
\begin{center}
\centerline{
\Qcircuit @C=1em @R=1em {
\lstick{\ket{q}}		& \qw		& \targ	& \targ & \targ & \qw & \qw & \qw & \qw	\\
\lstick{\ket{0}}		& \gate{H}	& \ctrl{-1} & \qw & \qw & \gate{H} & \ctrl{1} & \ctrl{2} & \qw \\
\lstick{\ket{0}}		& \gate{H}	& \qw & \ctrl{-2} & \qw & \gate{H} & \ctrl{2} & \qw & \ctrl{1} \\
\lstick{\ket{0}}		& \gate{H}	& \qw & \qw & \ctrl{-3} & \gate{H} & \qw & \ctrl{1} & \ctrl{1} \\
\lstick{\ket{a}=\ket{0}} & \qw		& \qw & \qw & \qw & \qw & \targ & \targ & \targ \\
}}
\caption[Strong Majority Vote]{Possible implementation of the strong majority vote from the initial qubit ($q$) to the ancilla ($a$).}\label{fig:strong_vote}
\end{center}
\end{figure}

\section{Fault-Tolerant Gates}

In order to calculate a threshold, we have to specify what set of gates we are going use. All the circuits which we construct must be in terms of these primitives. The list specified here is certainly not minimal, but the more gates we have, the simpler the circuits that we can construct.
\begin{enumerate}
\item Hadamard Gate, $H$
\item Bit-Flip, $X$
\item Phase-Flip, $Z$, and root, $\sqrt{Z}=S$
\item controlled-{\sc not}
\item {\sc swap}
\item $\pi/8$ gate, $T$
\item Toffoli gate (controlled-controlled-{\sc not})
\end{enumerate}

With this set of gates, we have to construct circuits on the next level of encoded qubits to implement the entire set, and a circuit to implement error correction of the encoded qubits (this circuit is referred to as {\sc ec}), such that if a single error occurs in any of the circuits, this affects no more than {\em{one}} of the qubits on each logical qubit.

The construction of some of these gates requires some additional sub-circuits including, in particular,
\begin{enumerate}
\item Preparation of cat states on physical qubits, $\ket{00\ldots 0}+\ket{11\ldots 1}$.
\item Conversion of logical qubit to a classical repetition code, denoted $N$. e.g.
$$
\alpha\ket{0_L}+\beta\ket{1_L}\rightarrow\alpha\ket{00\ldots 0}+\beta\ket{11\ldots 1}
$$
\item Preparation of $\ket{0_L}$
\item State preparation, particularly $\ket{0_L}+e^{i\pi/4}\ket{1_L}$, using an input state $\ket{0_L}$ \cite{Boykin:99}.
\item The error correction circuit, {\sc ec}.
\end{enumerate}

\subsection{Bitwise Gates}

The gates $H$, $X$, $Z$, $S$, c-{\sc not} and {\sc swap} are all applied bitwise, and are, therefore, comparatively simple. Their value of $A$ is almost entirely determined by the product of the number of applications of {\sc ec} and the number of locations in an {\sc ec}. There are slight differences if we assume we have a super-{\sc cu} such that all 7 gates can be applied simultaneously, or whether we have to apply them one at a time (in which case there are a number of `wait' operations, which contribute to the number of locations). However these are counted, the values of $A$ will not be comparable to some of the other gates.

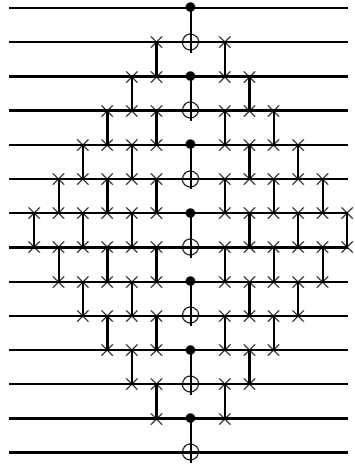
\begin{figure}
\begin{center}
\centerline{
\Qcircuit @C=1em @R=1em {
& \qw & \qw & \qw & \qw & \qw & \qw & \ctrl{1} & \qw & \qw & \qw & \qw & \qw & \qw \\
& \qw & \qw & \qw & \qw & \qw & \qswap & \targ & \qswap & \qw & \qw & \qw & \qw & \qw \\
& \qw & \qw & \qw & \qw & \qswap	& \qswap \qwx & \ctrl{1} & \qswap \qwx & \qswap & \qw & \qw & \qw & \qw \\
& \qw & \qw & \qw & \qswap	& \qswap \qwx	& \qswap & \targ & \qswap	& \qswap \qwx	& \qswap & \qw & \qw & \qw \\
& \qw & \qw & \qswap	& \qswap \qwx	& \qswap	& \qswap \qwx & \ctrl{1} & \qswap \qwx	& \qswap	& \qswap \qwx	& \qswap & \qw & \qw  \\
& \qw & \qswap	& \qswap \qwx	& \qswap	& \qswap \qwx	& \qswap & \targ & \qswap	& \qswap \qwx	& \qswap	& \qswap \qwx	& \qswap & \qw \\
& \qswap	& \qswap \qwx	& \qswap	& \qswap \qwx	& \qswap	& \qswap \qwx & \ctrl{1} & \qswap \qwx & \qswap	& \qswap \qwx	& \qswap	& \qswap \qwx	& \qswap  \\	
& \qswap \qwx	& \qswap	& \qswap \qwx	& \qswap	& \qswap \qwx	& \qswap & \targ & \qswap & \qswap \qwx	& \qswap	& \qswap \qwx	& \qswap	& \qswap \qwx   \\	
& \qw	& \qswap \qwx	& \qswap	& \qswap \qwx	& \qswap	& \qswap \qwx & \ctrl{1} & \qswap \qwx	& \qswap 	& \qswap \qwx	& \qswap	& \qswap \qwx & \qw  \\
& \qw	& \qw & \qswap \qwx	& \qswap	& \qswap \qwx	& \qswap & \targ & \qswap	& \qswap \qwx	& \qswap	& \qswap \qwx & \qw & \qw  \\
& \qw & \qw & \qw & \qswap \qwx	& \qswap	& \qswap \qwx & \ctrl{1} & \qswap \qwx	& \qswap	& \qswap \qwx & \qw & \qw & \qw  \\
& \qw & \qw & \qw & \qw & \qswap \qwx	& \qswap & \targ & \qswap & \qswap \qwx & \qw & \qw & \qw & \qw  \\
& \qw & \qw & \qw & \qw & \qw & \qswap \qwx & \ctrl{1} & \qswap \qwx & \qw & \qw & \qw & \qw & \qw \\
& \qw & \qw & \qw & \qw & \qw & \qw & \targ & \qw & \qw & \qw & \qw & \qw & \qw \\
}}
\caption[Encoded c{\sc not} gate with nearest-neighbour interactions]{Implementation of the encoded controlled-{\sc not} gate, where gates are required to be between nearest-neighbours.}\label{fig:cnot_swap}
\end{center}
\end{figure}

Let us demonstrate the required counting with an example of an encoded c{\sc not} gate. Consider Fig.~\ref{fig:EC_concept} ignoring, for now, the blocks of error correction. No two c{\sc not} gates act on the same qubits, so in principle all seven of them can be performed simultaneously. Each c{\sc not} gate counts as a single location, so this gate has 7 locations. Since this gate does not require any measurements, the result is the same when we remove the ability to perform measurements. If we remove the parallelism restriction, then only one gate can be applied at a time. Therefore, there are 13 locations for each of 7 time steps (12 locations where the `wait' operation happens, and a single location for the c{\sc not} that we are applying). This gives 91 locations. With a restriction to gates between nearest-neighbours, we must introduce a series of {\sc swap} gates to interlace the qubits from the two different logical qubits, as depicted in Fig.~\ref{fig:cnot_swap}. The {\sc swap} gates are not correctly ordered for the sake of space. However, it is clear that they can be implemented such that the {\sc cu} would only have to jump to its nearest-neighbour to be able to implement the next gate. This requires a total of 42 {\sc swap} gates, which are implemented one at a time. Hence, the total location count is $13\times(42+7)=637$. Finally, we move to the physical model, where we need to count the number of physical operations required to generate each of the interactions. In particular, to move the {\sc cu} from one qubit to its neighbour requires 3 {\sc swap} operations. Hence the total number of time steps increases by a factor of 4. The new location count is $13\times(42+7)+14\times 3\times(42+7)=2695$. However, this is actually somewhat misleading for a threshold argument because these additional {\sc swap} gates only affect the lowest level of concatenation. All higher levels just act on logical qubits, which do not have these separations. We can take this into account when calculating the threshold. Let $\epsilon^{(1)}_0$ be the threshold for the first level of concatenation and higher, while $\epsilon^{ph}_0$ is the threshold at the physical level. We can therefore write that
$$
\epsilon^{(1)}_0= B(\epsilon^{ph}_0)^2+\binom{A}{3}(\epsilon^{ph}_0)^3,
$$
which should improve our threshold estimate, $\epsilon^{ph}_0$.

\subsection{The $\pi/8$ Gate}

When constructing a universal gate set, one only needs either the Toffoli gate or the $\pi/8$ gate. The $\pi/8$ gate typically gives a lower threshold value due to the fact that it acts on fewer qubits. However, the Toffoli, as described below, is very useful to us because we perform so many of these gates during the coherent feedback part of error correction.

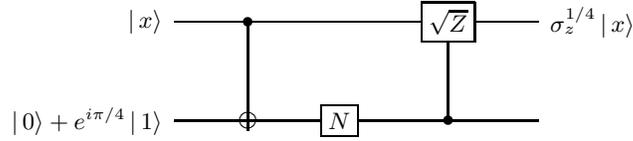
\begin{figure}
\begin{center}
\leavevmode
\centering
\Qcircuit{
\lstick{\ket{x}}					& \ctrl{1}	& \qw		& \gate{\sqrt{Z}}	& \rstick{\sigma_z^{1/4}\ket{x}} \qw		\\
\lstick{\ket{0}+e^{i\pi/4}\ket{1}}	& \targ		& \gate{N}	& \ctrl{-1} & \qw \\
}
\caption[Fault-tolerant T gate]{Fault-tolerant construction of the $\pi/8$ gate, T.}
\label{fig:T}
\end{center}
\end{figure}

In order to apply the $\pi/8$ gate, we first make an ancilla state, $\ket{0_L}+e^{i\pi/4}\ket{1_L}$, as detailed in \cite{Boykin:99}. This is input to the circuit specified in Fig. \ref{fig:T}, where the only additional component that we require is the gate $N$, which is specified below.

\subsection{The Toffoli Gate}

The Toffoli gate, or controlled-controlled-{\sc not}, is the worst-case gate, simply because of the number of inputs and outputs. The typical fault-tolerant construction was first suggested by Shor \cite{Shor_faults}, and has been altered to work deterministically in the absence of measurements \cite{Aharanov:99, Boykin:99}. We might also consider an alternative construction, using Fig.~\ref{fig:toffoli}. At first glance, this circuit would seem to be much worse because it includes 6 applications of $T$, and hence 6 applications of $N$, compared to only 2 applications in \cite{Boykin:99}. However, we may be able to gain some advantage by surrounding each gate by {\sc ec} circuits. In this case, all single errors are corrected within each gate. Hence the only pairs of locations which are not benign are contained within blocks of {\sc ec}-gate-{\sc ec}. Therefore, this could, potentially, represent a significant reduction in the number of non-benign pairs of locations, even if the number of locations has increased. Note that due to our available parallelism, we will also surround all the `wait' operations with {\sc ec}s as well.

We must remember that we are limited to nearest-neighbour interactions, which means that we must keep track of the locations of the ancilla qubits, and count the {\sc swap} operations required to move past these as well. It may also be useful to note that it does not matter which qubit the target qubit is, the number of required operations is the same (a {\sc swap} operation moves from the start of the circuit to the end).

It turns out that this operation is superior for a threshold under the most restrictive set of assumptions, which relates to our model of global control. Given that this represents an improvement, it is also relevant to ask whether it is more efficient to not include the Toffoli in our set of gates, and just expand it each time it is used in the {\sc ec} circuit. This would allow us to use a gate with fewer inputs and outputs as the worst-case gate, which could, potentially counter-act the increase in size of the {\sc ec} circuit.
\begin{figure}
\begin{center}
\leavevmode
\centering
\Qcircuit @C=1em @R=.7em {
	& \qw			& \qw		& \qw		& \qw		& \ctrl{1}	&\qw				& \qw		& \qw		& \ctrl{1}	& \qw				& \qw			& \ctrl{1}	& \qw				& \ctrl{1}	& \gate{T}	& \qw	\\
	& \qswap		& \gate{H}	& \targ		&\gate{T}	& \targ		&\gate{T^\dagger}	& \targ		&\gate{T}	& \targ		&\gate{T^\dagger}	& \qswap		&\targ		& \gate{T^\dagger}	& \targ		& \qw		& \qw	\\
	& \qswap \qwx	& \qw		& \ctrl{-1}	&\qw		& \qw		& \qw				& \ctrl{-1}	&\qw		& \qw		&\gate{T}			& \qswap \qwx	& \gate{H}	& \qw				& \qw		& \qw		& \qw	\\
}
\caption[Fault-tolerant Toffoli gate]{Toffoli gate, constructed out of other primitives at the same level of concatenation.}
\label{fig:toffoli}
\end{center}
\end{figure}
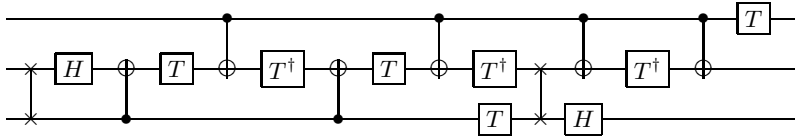

\begin{figure}
\begin{center}
\leavevmode
\centering
\Qcircuit @C=1em @R=.7em {
& \ctrl{3} & \qw & \qw & \qw & \qw & \qw & \gate{Z} & \ctrl{2} & \qw & \qw & \targ & \rstick{\ket{x}} \qw	\\
\lstick{\ket{AND} \Bigg\{} & \qw & \ctrl{3} & \qw & \qw & \qw & \qw & \ctrl{-1} & \qw & \targ & \ctrl{1} & \qw & \rstick{\ket{y}} \qw	\\
& \qw & \qw & \targ & \qw & \qw & \gate{Z} & \qw & \targ & \qw & \targ & \qw & \rstick{\ket{z\oplus(x\cdot y)}} \qw	\\
\lstick{\ket{x}} & \targ & \qw & \qw & \qw & \gate{N} & \qw & \qw & \qw & \qw & \ctrl{-1} & \ctrl{-3} &	\\
\lstick{\ket{y}} & \qw & \targ & \qw & \qw & \gate{N} & \qw & \qw & \ctrl{-2} & \ctrl{-3} & \qw & \qw &	\\
\lstick{\ket{z}} & \qw & \qw & \ctrl{-3} & \gate{H} & \gate{N} & \ctrl{-3} & \ctrl{-4} & \qw & \qw & \qw & \qw &	\\
}
\caption[Fault-tolerant Toffoli gate]{Alternative formulation of the Toffoli gate. The auxiliary state $\ket{AND}=\ket{000}+\ket{010}+\ket{100}+\ket{111}$ must also be fault-tolerantly constructed.}
\label{fig:toffoli2}
\end{center}
\end{figure}
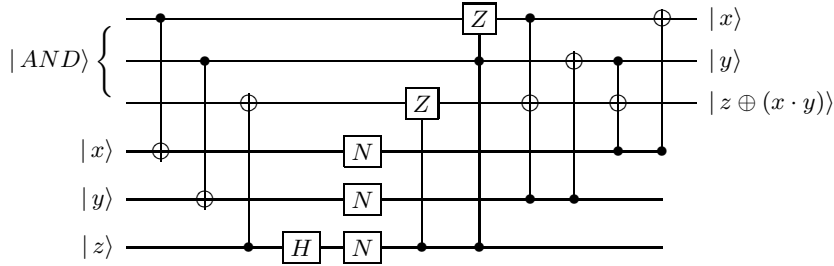

\subsection{N}

This is by far the most complicated gate, and follows the construction of \cite{Boykin:99}. As specified in that paper, the action of the gate is to take two inputs. One is a logical qubit, $\alpha\ket{0_L}+\beta\ket{1_L}$, and the other is a set of 7 qubits, all in the physical $\ket{0}$ state. The output is then of the form
$$
\alpha\ket{0_L}\ket{0000000}+\beta\ket{1_L}\ket{1111111}.
$$
However, we have chosen to represent this as a one-qubit gate to clearly indicate the fact that, in constructing this state, a single error can have a catastrophic effect on the logical states, and so they can no longer be used in the computation (only the physical qubits can).

The output of this gate is not a logical qubit, as with other gates. It is, instead, a classical repetition code. Given that we will never perform error correction on this circuit, this is perfectly allowable. Any gates that are controlled off this repetition code can be performed bitwise.

In essence, the gate is constructed by making use of the observation that the codewords of the Steane code have an even (odd) number of $\ket{1}$s for the $\ket{0_L}$ ($\ket{1_L}$) state. Hence, performing a weak majority vote from the codeword onto an ancilla gives one of the seven bits required for the classical repetition code. Since we must repeat this sequence seven times, controlled by the same qubits in the codeword, a strong majority vote should first be used. This would perhaps be the way that would give the smallest threshold (see Sec.~\ref{sec:benign}), although we chose to directly follow the circuits given in \cite{Boykin:99}, which replace the strong majority voting step with a stabilization process (syndrome extraction).

\section{Fault-Tolerant Threshold for the Computation}

\subsection{Circuit for {\sc ec}}

We have to be careful about the construction of a fault-tolerant error-correcting circuit. In particular, the error-correcting circuits previously shown, such as in Fig.~\ref{fig:steane} are not fault-tolerant, because if an error occurs on a qubit while the syndrome is being measured, the error remains on this qubit, and a different qubit is `corrected'. Instead, we need to create a different circuit, that protects against errors of this form by using a degree of redundancy, and then majority voting, as described in Sec.~\ref{sec:cat}.

\begin{figure}
\begin{center}
\resizebox*{17cm}{!}{
\Qcircuit @C=.5em @R=.7em {
	& \gate{\lightning} 	& \gate{H}	& \ctrl{3}	& \gate{H}	& \ctrl{13} & \qw 		& \qw 		& \qw 		& \qw 		& \targ		& \gate{\lightning}		& \gate{H}	& \ctrl{3}	& \gate{H}	& \ctrl{13} & \qw 		& \qw 		& \qw 		& \qw 		& \targ		& \gate{\lightning}		& \gate{H}	& \ctrl{3}	& \gate{H}	& \ctrl{13} & \qw 		& \qw 		& \qw 		& \qw 	 	& \targ 	& \qw 	\\
	& \gate{\lightning} 	& \qw		& \targ		& \gate{H}	& \qw 		& \ctrl{11} & \qw 		& \qw 		& \qw 	 	& \targ 	& \gate{\lightning} 	& \qw		& \targ		& \gate{H}	& \qw 		& \ctrl{10} & \qw 		& \qw 		& \qw 	 	& \targ 	& \gate{\lightning} 	& \qw		& \targ		& \gate{H}	& \qw 		& \ctrl{11} & \qw 		& \qw 		& \qw 	 	& \targ 	& \qw	\\
	& \gate{\lightning} 	& \qw		& \targ		& \gate{H}	& \qw 		& \qw 		& \ctrl{9} 	& \qw 		& \qw 	 	& \targ		& \gate{\lightning} 	& \qw		& \targ		& \gate{H}	& \qw 		& \qw 		& \ctrl{7}	& \qw 		& \qw 	 	& \targ		& \gate{\lightning} 	& \qw		& \targ		& \gate{H}	& \qw 		& \qw 		& \ctrl{7} 	& \qw 		& \qw 	 	& \targ 	& \qw	\\
	& \gate{\lightning} 	& \qw		& \targ		& \gate{H}	& \qw 		& \qw 		& \qw 		& \ctrl{7} 	& \qw 	 	& \targ		& \gate{\lightning} 	& \qw		& \targ		& \gate{H}	& \qw 		& \qw 		& \qw 		& \ctrl{4} 	& \qw 	 	& \targ		& \gate{\lightning} 	& \qw		& \targ		& \gate{H}	& \qw 		& \qw 		& \qw 		& \ctrl{5} 	& \qw 	 	& \targ 	& \qw	\\
	& \gate{\lightning} 	& \qw		& \qw		& \qw 		& \qw 		& \qw 		& \qw 		& \qw 		& \gate{H} 	& \ctrl{-4}	& \gate{H} 				& \qw		& \qw		& \qw 		& \qw 		& \qw 		& \qw 		& \qw 		& \qw 		& \qw 		& \qw 					& \qw		& \qw		& \qw 		& \qw 		& \qw 		& \qw 		& \qw 		& \qw 		& \qw 		& \qw 		& \ctrl{1} 	& \gate{X} 	& \ctrl{1} & \qw 		& \ctrl{1} 	& \gate{X}	& \ctrl{1}	& \qw 		& \ctrl{1}	& \qw 		& \ctrl{1} 	& \gate{X} 	& \ctrl{1} \\
	& \qw 					& \qw		& \qw		& \qw 		& \qw 		& \qw 		& \qw 		& \qw 		& \qw 		& \qw 		& \gate{\lightning} 	& \qw		& \qw		& \qw 		& \qw 		& \qw 		& \qw 		& \qw 		& \gate{H} 	& \ctrl{-5} & \gate{H} 				& \qw		& \qw		& \qw 		& \qw 		& \qw 		& \qw 		& \qw 		& \qw 		& \qw 		& \qw 		& \ctrl{1} 	& \qw 		& \ctrl{1} & \gate{X} 	& \ctrl{1} 	& \qw 		& \ctrl{1} 	& \qw 		& \ctrl{1} 	& \gate{X}	& \ctrl{1} 	& \qw		& \ctrl{1} \\
	& \qw 					& \qw		& \qw		& \qw 		& \qw 		& \qw 		& \qw 		& \qw 		& \qw 		& \qw 		& \qw 					& \qw		& \qw		& \qw 		& \qw 		& \qw 		& \qw 		& \qw 		& \qw 		& \qw 		& \gate{\lightning} 	& \qw		& \qw		& \qw 		& \qw 		& \qw 		& \qw 		& \qw 		& \gate{H} 	& \ctrl{-6} & \gate{H} 	& \ctrl{7}  & \qw 		& \ctrl{3} & \qw 		& \ctrl{2} 	& \qw 		& \ctrl{6} 	& \gate{X}	& \ctrl{4} 	& \qw 		& \ctrl{5} 	& \qw	 	& \ctrl{1}	\\
	& \qw 					& \qw		& \qw		& \qw 		& \qw 		& \qw 		& \qw 		& \qw 		& \qw 		& \qw 		& \qw 					& \qw		& \qw		& \qw 		& \qw 		& \qw 		& \qw 		& \targ 	& \qw 		& \qw 		& \qw 					& \qw		& \qw		& \qw 		& \qw 		& \qw 		& \qw 		& \qw 		& \gate{H} 	& \qw 		& \qw 		& \qw 		& \qw		& \qw		& \qw 		& \qw 		& \qw 		& \qw 		& \qw 		& \qw 		& \qw 		& \qw 		& \qw		& \targ 	\\
	& \qw 					& \qw		& \qw		& \qw 		& \qw 		& \qw 		& \qw 		& \qw 		& \qw 		& \qw 		& \qw 					& \qw		& \qw		& \qw 		& \qw 		& \qw 		& \qw 		& \qw 		& \qw 		& \qw 		& \qw 					& \qw		& \qw		& \qw 		& \qw 		& \qw 		& \qw 		& \targ 	& \gate{H} 	& \qw 		& \qw 		& \qw 		& \qw		& \qw		& \qw 		& \targ 	& \qw 		& \qw 		& \qw 		& \qw 		& \qw 		& \qw 		& \qw		& \qw 	\\
	& \qw 					& \qw		& \qw		& \qw 		& \qw 		& \qw 		& \qw 		& \qw 		& \qw 		& \qw 		& \qw 					& \qw		& \qw		& \qw 		& \qw 		& \qw 		& \targ 	& \qw 		& \qw 		& \qw 		& \qw 					& \qw		& \qw		& \qw 		& \qw 		& \qw 		& \targ 	& \qw 		& \gate{H} 	& \qw 		& \qw 		& \qw 		& \qw		& \targ		& \qw 		& \qw 		& \qw 		& \qw 		& \qw 		& \qw 		& \qw 		& \qw 		& \qw		& \qw 	\\
	& \qw 					& \qw		& \qw		& \qw 		& \qw 		& \qw 		& \qw 		& \targ 	& \qw 		& \qw 		& \qw 					& \qw		& \qw		& \qw 		& \qw 		& \qw 		& \qw 		& \qw 		& \qw 		& \qw 		& \qw 					& \qw		& \qw		& \qw 		& \qw 		& \qw 		& \qw 		& \qw 		& \gate{H} 	& \qw 		& \qw 		& \qw 		& \qw		& \qw		& \qw 		& \qw 		& \qw 		& \qw 		& \qw 		& \targ 	& \qw 		& \qw 		& \qw		& \qw 	\\
	& \qw 					& \qw		& \qw		& \qw 		& \qw 		& \qw 		& \targ 	& \qw 		& \qw 		& \qw 		& \qw 					& \qw		& \qw		& \qw 		& \qw 		& \targ 	& \qw 		& \qw 		& \qw 		& \qw 		& \qw 					& \qw		& \qw		& \qw 		& \qw 		& \qw 		& \qw 		& \qw 		& \gate{H} 	& \qw 		& \qw 		& \qw 		& \qw		& \qw		& \qw 		& \qw 		& \qw 		& \qw 		& \qw 		& \qw 		& \qw 		& \targ 	& \qw		& \qw 	\\
	& \qw 					& \qw		& \qw		& \qw 		& \qw 		& \targ 	& \qw 		& \qw 		& \qw 		& \qw 		& \qw 					& \qw		& \qw		& \qw 		& \qw 		& \qw 		& \qw 		& \qw 		& \qw 		& \qw 		& \qw 					& \qw		& \qw		& \qw 		& \qw 		& \targ 	& \qw 		& \qw 		& \gate{H} 	& \qw 		& \qw 		& \qw 		& \qw		& \qw		& \qw 		& \qw 		& \qw 		& \targ		& \qw 		& \qw 		& \qw 		& \qw 		& \qw		& \qw 	\\
	& \qw 					& \qw		& \qw		& \qw 		& \targ 	& \qw 		& \qw 		& \qw 		& \qw 		& \qw 		& \qw 					& \qw		& \qw		& \qw 		& \targ 	& \qw 		& \qw 		& \qw 		& \qw 		& \qw 		& \qw 					& \qw		& \qw		& \qw 		& \targ 	& \qw 		& \qw 		& \qw 		& \gate{H} 	& \qw 		& \qw 		& \targ		& \qw		& \qw		& \qw 		& \qw 		& \qw 		& \qw 		& \qw 		& \qw 		& \qw 		& \qw 		& \qw		& \qw 		
}
}
\caption[Fault-tolerant error correcting circuit]{The {\sc ec} circuit, designed such that a single error during the error correction process does not affect more than one of the output qubits. The circuit shown only corrects for $Z$ errors. To correct for $X$ errors, the circuit needs to be repeated.} \label{fig:FT_EC}
\end{center}
\end{figure}
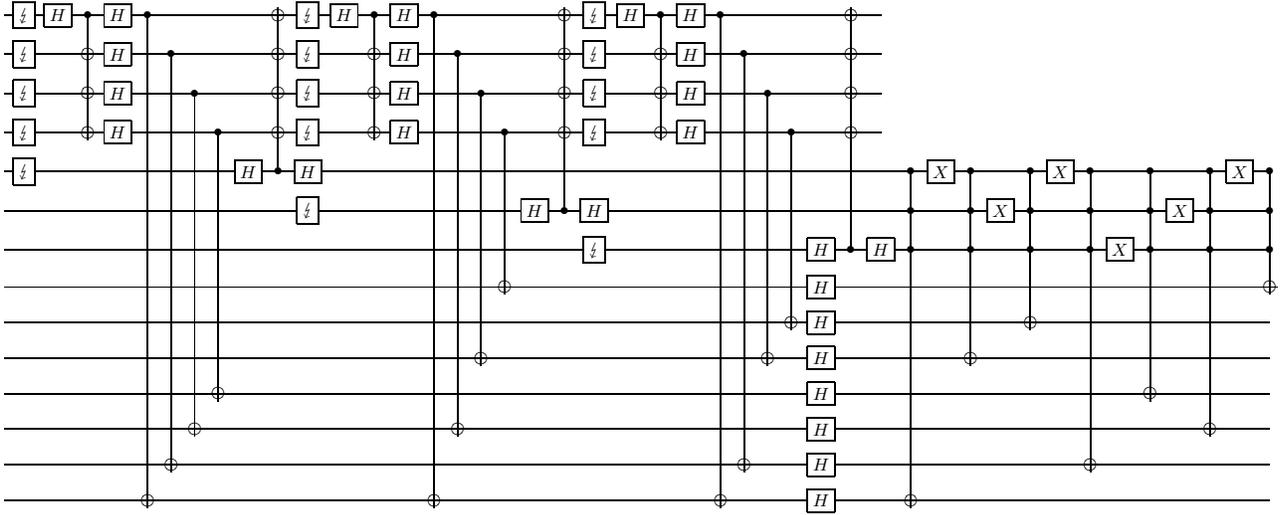

In terms of the primitives defined, the circuit in Fig.~\ref{fig:FT_EC} fulfills all the requirements of a circuit that corrects single errors on its inputs, and does not catastrophically propagate errors that occur during the circuit. Depending on how much parallelism is available (i.e.~whether we can perform more than one operation simultaneously on a given block), the arrangement of ancillas (and in particular, the number of ancillas used) can be optimised to reduce the number of locations in the circuit.

Note that a single application of this circuit for error correction (while we require two applications for error correction) can also be used to fault-tolerantly prepare an encoded qubit in the $\ket{0_L}$ state, just by supplying $\ket{0000000}$ in place of the logical qubit to be corrected. Given that we only need to prepare the $\ket{0_L}$ state at the beginning of any of our gates, we can incorporate this part in the application of the {\sc ec} before the start of each gate, thereby reducing the required number of time steps.

For simplicity, we have left one important part out of the circuit in Fig.~\ref{fig:FT_EC}, which corrects a significant oversight in the circuit as shown. In particular, if a single error occurs on one of the computational qubits between sets of controlled-{\sc not} gates that feed the error information onto the qubits, then not only do we have a fault on that qubit, but we apply a correction to a different qubit. This is avoided at the start of the circuit by performing a strong majority vote process on each of the computational qubits which is to be read more than once. In the following subsection, we will learn that we are further justified in not depicting this part of the circuit because the extra gates are benign with respect to the rest of the circuit, and hence makes a negligible difference to the threshold estimate that we will make (however, we must remember that this process takes a certain number of steps, and adds to the number of error locations on the rest of the circuit).

With this circuit in place, we are now in a position to enumerate the number of locations for each type of gate. These results are given in Tab.~\ref{tab:results}. We therefore take the Toffoli gate as being the one with the most locations (once we've placed error correcting units on each input and output). In this case $A=447357+6\times 19724=565701$, and an approximate error threshold is given by $\binom{A}{2}^{-1}=6.2\times 10^{-12}$.
\begin{table}[!t]
\begin{center}
\centerline{
\begin{tabular}{|c|c|c|c|c|c|c|c|}
\hline
Gate & Inputs & No & No & Restricted & NN & Physical	\\
& \& Outputs & restrictions & measurements & parallelism & interactions & Model \\
\hline
$H$, $S$, $Z$, $X$ & 1 & 7 & 7 & 49 & 49 & 196	\\
c{\sc not} & 2 & 7 & 7 & 91 & 637 & 2695	\\
T & 1 & 237 & 2023 & 4926 & 10498 & 34018 \\
Toffoli & 3 & N/A & 6279 & 41952 & 123663 & 447357	\\
$N$ & 1\footnotemark & N/A & 1127 & 3078 & 5990 & 15692 \\
{\sc ec} & N/A & 142 & 1486 & 2928 & 4982 & 19724 \\
\hline
\end{tabular}
}
\caption[Number of error locations for encoded gates]{Number of locations for each gate type. As the columns go from left to right, we add more restrictions leading, finally, to the global control model.}\label{tab:results}
\end{center}
\end{table}
\footnotetext{This output is a classical repetition code of 7 qubits, not an encoded qubit.}

\subsection{Benign Locations} \label{sec:benign}

In \cite{gottesman:2005}, counting the benign locations for pairs of errors provided a significant enhancement to the fault-tolerant threshold. However, the task was far simpler in that paper because the gate constructions used far fewer operations (due to the availability of measurement, arbitrary parallelism and non-nearest-neighbour operations), and because the gates were constructed out of Clifford operators, which means that the effect of errors can easily be calculated by propagating the errors through the circuit, which can be efficiently simulated on a classical computer. As such, our task is far harder, and we do not intend to count all the benign pairs. However, we can make (or re-use from \cite{gottesman:2005}) some very simple arguments.
\begin{enumerate}
\item If there is a `wait' operation, then the locations either side of it form a pair. If there are multiple wait operations in a row, then all possible pairs of locations are benign.
\item This can be generalised because the circuits are constructed in such a way that if an error affects one qubit of the 7, then for all time it only affects this one, and its equivalent qubits in the other logical qubits of the gate. Hence, if a second error occurs on this qubit, it acts as only a single error, and, therefore, all pairs along these lines are benign.
\item All pairs of locations in an {\sc ec} that acts on the input qubits are benign \cite{gottesman:2005}.
\item Pairs of errors that occur on two different output {\sc ec} blocks are benign. This is clear because there are no gates that can cause the two errors to be present on the same logical qubit.
\item A single error that occurs within a strong majority vote (except for those on the output ancilla) is benign with respect to any single error that occurs externally to that process. This is because the majority vote clears the effect of that error, just leaving the single error which gets corrected. By using a cat state of at least 6 qubits, it can also be made benign with respect to pairs of errors that occur within the circuit.
\end{enumerate}

\subsection{Threshold}

We are now in a position to enumerate the number of operations required for each gate, and make a first estimate as to the number of benign pairs. We have chosen to split this into a series of steps, building up slowly to the final threshold. This allows us to see where the basic costs of a global control scheme come into effect. Firstly, we evaluate the threshold allowing arbitrary parallelism and measurements. For this, we get a value of $8.9\times 10^{-6}$. This is comparable to the value of $2.7\times 10^{-5}$, obtained by \cite{gottesman:2005}, but indicates that our simple evaluation of the number of benign pairs actually misses a large fraction. This suggests that all further results would benefit from more rigorous accounting of the benign pairs.

If we disallow measurements, then not only do our gate sequences get larger, but we also have to include the Toffoli gate in our set of gates. As a result, the threshold sees its most significant hit at this point, where we find $\epsilon_0=6.4\times 10^{-9}$. If we further assume that we have only a single {\sc cu} in each error correcting block, then all gates have to be performed sequentially, instead of in parallel, thereby adding a lot of extra `wait' operations. These further reduce the threshold to $\epsilon_0=8.1\times 10^{-10}$. Next, we must add in sets of {\sc swap} operations so that the gates are always performed between nearest-neighbours and, further, even single-qubit gates must be performed on adjacent qubits. These leave us with a threshold of $\epsilon_0=2.6\times 10^{-10}$. Recall that adding in these nearest-neighbour interactions would mean that the circuits are not fault-tolerant if we were to apply the {\sc swap}s (although it is known how to make such a structure fault-tolerant \cite{boykin-2006}). However, in the present case, these gates just serve to count the number of operations while the {\sc cu} is moving independently. Hence, this calculation is valid here, but not for a general nearest-neighbour scheme.

Finally, we must take into account the physical model where we actually have to perform 3 {\sc swap} operations to move between adjacent computational qubits. As already discussed, however, this only has to be done at the lowest level of concatenation, which means that its contribution is not as significant as it might otherwise have been. The final threshold that we find is
$$
\epsilon_0=6.8\times 10^{-11}
$$
This threshold simply provides a bound -- the real threshold is certainly higher. However, the intention of this calculation was not to optimise this bound, simply to show that a bound exists for a global control scheme. In particular, theoretical calculations of thresholds, such as \cite{gottesman:2005, Aharanov:99}, give results several orders of magnitude worse than current numerical computations indicate. For example, while \cite{gottesman:2005} calculates a threshold of $2.7\times 10^{-5}$, numerical estimates \cite{Knill:04} put the threshold for similar assumptions to be closer to 0.03. We estimate that, had we used the strong majority voting version of the gate $N$, instead of the version of \cite{Boykin:99}, the threshold would have been approximately $10^{-10}$, although making an accurate count of the number of benign locations is more demanding.

\section{Fault-Tolerant Threshold for the Classical States} \label{sec:classical}

To finish the argument, we must now demonstrate fault-tolerant error correcting circuits for the classical states and calculate their threshold, $\epsilon_c$. We expect such a threshold to be significantly larger because the encoding is much smaller (only 3 qubits as opposed to 7), and because our {\sc ec}s only have to correct for one type of error -- bit-flips. Phase-flips don't affect the final result.  Although only 3 {\sc cu}s may be active (per block), we have to remember that, actually, we are trying to stabilise all 7 of the {\sc cu}s associated with that level of concatenation.

As before, we must construct a list of the required operations, and build them out of the circuit primitives. In this case, the most costly operation will involve moving a {\sc cu}'s resting place from one {\sc ss} to its neighbour. At all levels of concatenation except the lowest, the move operation involves four controlled-{\sc not} gates (two in each {\sc ss}) and seven {\sc swap} gates (to move the {\sc cu} from one {\sc ss} to the other). However, in our previous calculation of the threshold, we did not decompose the two-qubit gate into the one-qubit gate steps, we merely took it as a primitive. The {\sc cu}'s manipulation of {\sc ss}s is exactly the same as a two-qubit gate, and so we shall count each of these as a primitive. At the lowest level of concatenation, we must also move past all the buffer qubits, requiring 28 {\sc swap} operations instead.

Before and after this move operation, we must apply error correction to the {\sc cu}s. This involves changing the array of {\sc cu}s available (one step), and then performing the procedures described in Sec.~\ref{sec:ec_CU}. These procedures correct one of the three {\sc cu}s, and hence must be repeated three times to ensure that errors do not propagate catastrophically. Subsequently, we use these 3 to reset all the {\sc cu}s from the previous level of concatenation. This step is necessary because we are not operating, in the end, on encoded qubits, but with the single {\sc cu}s and must therefore propagate the extra stability due to this round of error correction to all the {\sc cu}s. As a result, we have a total of $\sim 1800$ locations at which errors can occur. This gives $1.6\times 10^{6}$ pairs of locations at which faults could occur. Making no effort to enumerate the benign pairs, we simply quote $\epsilon_c=6\times 10^{-7}$, realising that this is insignificant compared to the value of $7\times 10^{-11}$, above, as we expected. If we have a physical implementation in which we can match the main threshold, $\epsilon_0$, and the classical states obey a threshold $\epsilon_c$, then we can implement approximately
$$
\left(\frac{\epsilon_c}{\epsilon_0}\right)^2
$$
operations between each phase of error correction on the {\sc cu}s. Hence, the classical threshold will have an insignificant contribution to the overall threshold, justifying our exclusion of it from the full calculation. We have skipped over arguing that the operations we perform uphold the requirements of non-propagation of single errors. One could most simply justify that this can be done by moving to a 5-bit code instead of a 3-bit code. That way, even if a single error occurs during the error correction step, and discounting the particular bit being corrected, there are still 3 other unaffected bits forming a majority, and hence single errors can be prevented from propagating.

\section{Summary and Conclusions}

The main result of this paper is simply stated -- that a globally controlled architecture has a fault-tolerant threshold which is a positive number. In achieving this result, we have made two basic assumptions. Firstly, we have treated the classical and quantum states as two distinct sections, and only errors on the computational qubits contribute to the final threshold. We have justified this by also calculating a threshold for the classical bits, which is much smaller. However, a more elegant approach would be to combine the two elements. The second, implicit, assumption is that our computer was correctly initialised with the required patterning of classical states. One might expect that our fault-tolerant protocols would enable us to correctly initialise these patterns from some smaller initial configuration but there is, as yet, no rigour behind these expectations.

Given that little effort was expended in optimising the threshold, one might expect that significant improvements can be made to the calculated value. One is also given hope, since it was observed in \cite{kay-2006-73} that moving to a global control scheme allows alterations to (in this case) typical optical lattice schemes, such as changing from red-detuned lasers to blue-detuned, which brings significant (order of magnitude) benefits to some decoherence mechanisms, thereby compensating for some of the cost of moving to such a scheme.

An open question that still remains is whether, for other global control schemes, the basis states (such as those required for use with cellular automata), can be stabilised. Our results only apply at the level of logical qubits, not the underlying physical model, except that the two coincide for our chosen model.

This work was supported by Clare College, Cambridge and the European
Commission through the Integrated Projects SCALA (CT-015714) and QAP (IST-3-015848).


\begin{thebibliography}{31}
\expandafter\ifx\csname natexlab\endcsname\relax\def\natexlab#1{#1}\fi
\expandafter\ifx\csname bibnamefont\endcsname\relax
  \def\bibnamefont#1{#1}\fi
\expandafter\ifx\csname bibfnamefont\endcsname\relax
  \def\bibfnamefont#1{#1}\fi
\expandafter\ifx\csname citenamefont\endcsname\relax
  \def\citenamefont#1{#1}\fi
\expandafter\ifx\csname url\endcsname\relax
  \def\url#1{\texttt{#1}}\fi
\expandafter\ifx\csname urlprefix\endcsname\relax\def\urlprefix{URL }\fi
\providecommand{\bibinfo}[2]{#2}
\providecommand{\eprint}[2][]{\url{#2}}

\bibitem[{\citenamefont{Benjamin}(2002)}]{Benjamin:2002a}
\bibinfo{author}{\bibfnamefont{S.~C.} \bibnamefont{Benjamin}},
  \bibinfo{journal}{Phys. Rev. Lett.} \textbf{\bibinfo{volume}{88}},
  \bibinfo{pages}{017904} (\bibinfo{year}{2002}).

\bibitem[{\citenamefont{Benjamin et~al.}(2003)\citenamefont{Benjamin, Bririd,
  and Kay}}]{Benjamin:2003b}
\bibinfo{author}{\bibfnamefont{S.~C.} \bibnamefont{Benjamin}},
  \bibinfo{author}{\bibfnamefont{A.}~\bibnamefont{Bririd}}, \bibnamefont{and}
  \bibinfo{author}{\bibfnamefont{A.}~\bibnamefont{Kay}} (\bibinfo{year}{2003}), \bibinfo{note}{quant-ph/0308113}.

\bibitem[{\citenamefont{Kay}(2005)}]{Kay:2005c}
\bibinfo{author}{\bibfnamefont{A.}~\bibnamefont{Kay}} (\bibinfo{year}{2005}),
  \bibinfo{note}{quant-ph/0504197}.

\bibitem[{\citenamefont{Kay}(2006{\natexlab{a}})}]{Kay:thesis}
\bibinfo{author}{\bibfnamefont{A.}~\bibnamefont{Kay}}, Ph.D. thesis,
  \bibinfo{address}{University of Cambridge}
  (\bibinfo{year}{2006}{\natexlab{a}}),
  \urlprefix\url{http://cam.qubit.org/users/Alastair/thesis.pdf}.

\bibitem[{\citenamefont{Christandl et~al.}(2004)\citenamefont{Christandl,
  Datta, Ekert, and Landahl}}]{Christandl}
\bibinfo{author}{\bibfnamefont{M.}~\bibnamefont{Christandl}},
  \bibinfo{author}{\bibfnamefont{N.}~\bibnamefont{Datta}},
  \bibinfo{author}{\bibfnamefont{A.}~\bibnamefont{Ekert}}, \bibnamefont{and}
  \bibinfo{author}{\bibfnamefont{A.~J.} \bibnamefont{Landahl}},
  \bibinfo{journal}{Phys. Rev. Lett.} \textbf{\bibinfo{volume}{92}},
  \bibinfo{pages}{187902} (\bibinfo{year}{2004}).

\bibitem[{\citenamefont{Christandl et~al.}(2005)\citenamefont{Christandl,
  Datta, Dorlas, Ekert, Kay, and Landahl}}]{Kay:2004c}
\bibinfo{author}{\bibfnamefont{M.}~\bibnamefont{Christandl}},
  \bibinfo{author}{\bibfnamefont{N.}~\bibnamefont{Datta}},
  \bibinfo{author}{\bibfnamefont{T.}~\bibnamefont{Dorlas}},
  \bibinfo{author}{\bibfnamefont{A.}~\bibnamefont{Ekert}},
  \bibinfo{author}{\bibfnamefont{A.}~\bibnamefont{Kay}}, \bibnamefont{and}
  \bibinfo{author}{\bibfnamefont{A.~J.} \bibnamefont{Landahl}},
  \bibinfo{journal}{Phys. Rev. A} \textbf{\bibinfo{volume}{71}},
  \bibinfo{pages}{032312} (\bibinfo{year}{2005}).

\bibitem[{\citenamefont{Kay}(2006{\natexlab{b}})}]{kay:2006a}
\bibinfo{author}{\bibfnamefont{A.}~\bibnamefont{Kay}},
  \bibinfo{journal}{Physical Review A} \textbf{\bibinfo{volume}{73}},
  \bibinfo{pages}{032306} (\bibinfo{year}{2006}{\natexlab{b}}).

\bibitem[{\citenamefont{Benjamin}(2004)}]{Benjamin:04}
\bibinfo{author}{\bibfnamefont{S.~C.} \bibnamefont{Benjamin}},
  \bibinfo{journal}{New J. Phys.} \textbf{\bibinfo{volume}{6}},
  \bibinfo{pages}{61} (\bibinfo{year}{2004}).

\bibitem[{\citenamefont{Lloyd}(1993)}]{Lloyd:1993a}
\bibinfo{author}{\bibfnamefont{S.}~\bibnamefont{Lloyd}},
  \bibinfo{journal}{Science} \textbf{\bibinfo{volume}{261}},
  \bibinfo{pages}{1569} (\bibinfo{year}{1993}).

\bibitem[{\citenamefont{Nielsen and Chuang}(2000)}]{nielsen}
\bibinfo{author}{\bibfnamefont{M.~A.} \bibnamefont{Nielsen}} \bibnamefont{and}
  \bibinfo{author}{\bibfnamefont{I.~L.} \bibnamefont{Chuang}},
  \emph{\bibinfo{title}{Quantum Computation and Quantum Information}}
  (\bibinfo{publisher}{Cambridge University Press},
  \bibinfo{address}{Cambridge, UK}, \bibinfo{year}{2000}).

\bibitem[{\citenamefont{Vollbrecht et~al.}(2004)\citenamefont{Vollbrecht,
  Solano, and Cirac}}]{Vollbrecht}
\bibinfo{author}{\bibfnamefont{K.}~\bibnamefont{Vollbrecht}},
  \bibinfo{author}{\bibfnamefont{E.}~\bibnamefont{Solano}}, \bibnamefont{and}
  \bibinfo{author}{\bibfnamefont{J.~I.} \bibnamefont{Cirac}},
  \bibinfo{journal}{Phys. Rev. Lett.} \textbf{\bibinfo{volume}{93}},
  \bibinfo{pages}{220502} (\bibinfo{year}{2004}).

\bibitem[{\citenamefont{Kay and Pachos}(2004)}]{kay-2004-6}
\bibinfo{author}{\bibfnamefont{A.}~\bibnamefont{Kay}} \bibnamefont{and}
  \bibinfo{author}{\bibfnamefont{J.~K.} \bibnamefont{Pachos}},
  \bibinfo{journal}{New Journal of Physics} \textbf{\bibinfo{volume}{6}},
  \bibinfo{pages}{126} (\bibinfo{year}{2004}).

\bibitem[{\citenamefont{Kay et~al.}(2006)\citenamefont{Kay, Pachos, and
  Adams}}]{kay-2006-73}
\bibinfo{author}{\bibfnamefont{A.}~\bibnamefont{Kay}},
  \bibinfo{author}{\bibfnamefont{J.~K.} \bibnamefont{Pachos}},
  \bibnamefont{and} \bibinfo{author}{\bibfnamefont{C.~S.} \bibnamefont{Adams}},
  \bibinfo{journal}{Physical Review A} \textbf{\bibinfo{volume}{73}},
  \bibinfo{pages}{022310} (\bibinfo{year}{2006}).

\bibitem[{\citenamefont{Calarco et~al.}(2004)\citenamefont{Calarco, Dorner,
  Julienne, Williams, and Zoller}}]{Zoller:1}
\bibinfo{author}{\bibfnamefont{T.}~\bibnamefont{Calarco}},
  \bibinfo{author}{\bibfnamefont{U.}~\bibnamefont{Dorner}},
  \bibinfo{author}{\bibfnamefont{P.}~\bibnamefont{Julienne}},
  \bibinfo{author}{\bibfnamefont{C.}~\bibnamefont{Williams}}, \bibnamefont{and}
  \bibinfo{author}{\bibfnamefont{P.}~\bibnamefont{Zoller}},
  \bibinfo{journal}{Phys. Rev. A} \textbf{\bibinfo{volume}{70}},
  \bibinfo{pages}{012306} (\bibinfo{year}{2004}).

\bibitem[{\citenamefont{Aharonov and Ben-Or}(1999)}]{Aharanov:99}
\bibinfo{author}{\bibfnamefont{D.}~\bibnamefont{Aharonov}} \bibnamefont{and}
  \bibinfo{author}{\bibfnamefont{M.}~\bibnamefont{Ben-Or}}
  (\bibinfo{year}{1999}), \bibinfo{note}{quant-ph/9906129}.

\bibitem[{\citenamefont{Boykin et~al.}(2002)\citenamefont{Boykin, Mor,
  Roychowdhury, Vatan, and Vrijen}}]{algo_cool}
\bibinfo{author}{\bibfnamefont{P.~O.} \bibnamefont{Boykin}},
  \bibinfo{author}{\bibfnamefont{T.}~\bibnamefont{Mor}},
  \bibinfo{author}{\bibfnamefont{V.}~\bibnamefont{Roychowdhury}},
  \bibinfo{author}{\bibfnamefont{F.}~\bibnamefont{Vatan}}, \bibnamefont{and}
  \bibinfo{author}{\bibfnamefont{R.}~\bibnamefont{Vrijen}},
  \bibinfo{journal}{Proc. Natl. Acad. Sci. USA} \textbf{\bibinfo{volume}{99}},
  \bibinfo{pages}{3388} (\bibinfo{year}{2002}).

\bibitem[{\citenamefont{Benjamin}(2000)}]{benjamin:2002b}
\bibinfo{author}{\bibfnamefont{S.~C.} \bibnamefont{Benjamin}},
  \bibinfo{journal}{Phys. Rev. A} \textbf{\bibinfo{volume}{61}},
  \bibinfo{pages}{020301} (\bibinfo{year}{2000}).

\bibitem[{\citenamefont{Benjamin}(2001)}]{Benjamin:2001b}
\bibinfo{author}{\bibfnamefont{S.}~\bibnamefont{Benjamin}}
  (\bibinfo{year}{2001}), \bibinfo{note}{quant-ph/0104117}.

\bibitem[{\citenamefont{Benjamin and Bose}(2003)}]{BenjaminBose1}
\bibinfo{author}{\bibfnamefont{S.~C.} \bibnamefont{Benjamin}} \bibnamefont{and}
  \bibinfo{author}{\bibfnamefont{S.}~\bibnamefont{Bose}},
  \bibinfo{journal}{Phys. Rev. Lett.} \textbf{\bibinfo{volume}{90}},
  \bibinfo{pages}{247901} (\bibinfo{year}{2003}).

\bibitem[{\citenamefont{Perez-Delgado and Cheung}(2005)}]{QCA}
\bibinfo{author}{\bibfnamefont{C.~A.} \bibnamefont{Perez-Delgado}}
  \bibnamefont{and} \bibinfo{author}{\bibfnamefont{D.}~\bibnamefont{Cheung}}
  (\bibinfo{year}{2005}), \bibinfo{note}{quant-ph/0508164}.

\bibitem[{\citenamefont{Lidar et~al.}(1999)\citenamefont{Lidar, Bacon, and
  Whaley}}]{Bacon:99}
\bibinfo{author}{\bibfnamefont{D.~A.} \bibnamefont{Lidar}},
  \bibinfo{author}{\bibfnamefont{D.}~\bibnamefont{Bacon}}, \bibnamefont{and}
  \bibinfo{author}{\bibfnamefont{K.~B.} \bibnamefont{Whaley}},
  \bibinfo{journal}{Phys. Rev. Lett.} \textbf{\bibinfo{volume}{82}},
  \bibinfo{pages}{4556} (\bibinfo{year}{1999}).

\bibitem[{\citenamefont{Bacon et~al.}(2000)\citenamefont{Bacon, Kempe, Lidar,
  and Whaley}}]{Bacon:00}
\bibinfo{author}{\bibfnamefont{D.}~\bibnamefont{Bacon}},
  \bibinfo{author}{\bibfnamefont{J.}~\bibnamefont{Kempe}},
  \bibinfo{author}{\bibfnamefont{D.~A.} \bibnamefont{Lidar}}, \bibnamefont{and}
  \bibinfo{author}{\bibfnamefont{K.~B.} \bibnamefont{Whaley}},
  \bibinfo{journal}{Phys. Rev. Lett.} \textbf{\bibinfo{volume}{85}},
  \bibinfo{pages}{1758} (\bibinfo{year}{2000}).

\bibitem[{\citenamefont{Wu et~al.}(2002)\citenamefont{Wu, Byrd, and
  Lidar}}]{Lidar:02}
\bibinfo{author}{\bibfnamefont{L.-A.} \bibnamefont{Wu}},
  \bibinfo{author}{\bibfnamefont{M.~S.} \bibnamefont{Byrd}}, \bibnamefont{and}
  \bibinfo{author}{\bibfnamefont{D.~A.} \bibnamefont{Lidar}},
  \bibinfo{journal}{Phys. Rev. Lett.} \textbf{\bibinfo{volume}{89}},
  \bibinfo{pages}{127901} (\bibinfo{year}{2002}).

\bibitem[{\citenamefont{Byrd et~al.}(2005)\citenamefont{Byrd, Lidar, Wu, and
  Zanardi}}]{Lidar:04}
\bibinfo{author}{\bibfnamefont{M.~S.} \bibnamefont{Byrd}},
  \bibinfo{author}{\bibfnamefont{D.~A.} \bibnamefont{Lidar}},
  \bibinfo{author}{\bibfnamefont{L.-A.} \bibnamefont{Wu}}, \bibnamefont{and}
  \bibinfo{author}{\bibfnamefont{P.}~\bibnamefont{Zanardi}},
  \bibinfo{journal}{Phys. Rev. A} \textbf{\bibinfo{volume}{71}},
  \bibinfo{pages}{052301} (\bibinfo{year}{2005}).

\bibitem[{\citenamefont{Mohseni and Lidar}(2005)}]{mohseni:05}
\bibinfo{author}{\bibfnamefont{M.}~\bibnamefont{Mohseni}} \bibnamefont{and}
  \bibinfo{author}{\bibfnamefont{D.~A.}~\bibnamefont{Lidar}},
  \bibinfo{journal}{Phys. Rev. Lett.} \textbf{\bibinfo{volume}{94}},
  \bibinfo{pages}{040507} (\bibinfo{year}{2005}).

\bibitem[{\citenamefont{Raussendorf}(2005)}]{raussendorf:05}
\bibinfo{author}{\bibfnamefont{R.}~\bibnamefont{Raussendorf}},
  \bibinfo{journal}{Phys. Rev. A} \textbf{\bibinfo{volume}{72}},
  \bibinfo{pages}{052301} (\bibinfo{year}{2005}).

\bibitem[{\citenamefont{Fitzsimons and Twamley}(2006)}]{fitz:06}
\bibinfo{author}{\bibfnamefont{J.}~\bibnamefont{Fitzsimons}} \bibnamefont{and}
  \bibinfo{author}{\bibfnamefont{J.}~\bibnamefont{Twamley}},
 \bibinfo{journal}{Phys. Rev. Lett.} \textbf{\bibinfo{volume}{97}},
  \bibinfo{pages}{090502} (\bibinfo{year}{2006}).

\bibitem[{\citenamefont{Aliferis et~al.}(2006)\citenamefont{Aliferis,
  Gottesman, and Preskill}}]{gottesman:2005}
\bibinfo{author}{\bibfnamefont{P.}~\bibnamefont{Aliferis}},
  \bibinfo{author}{\bibfnamefont{D.}~\bibnamefont{Gottesman}},
  \bibnamefont{and} \bibinfo{author}{\bibfnamefont{J.}~\bibnamefont{Preskill}},
  \bibinfo{journal}{Quant. Inf. Comput.} \textbf{\bibinfo{volume}{6}},
  \bibinfo{pages}{97} (\bibinfo{year}{2006}).

\bibitem[{\citenamefont{Boykin et~al.}(1999)\citenamefont{Boykin, Mor,
  Roychowdhury, and Vatan}}]{Boykin:99}
\bibinfo{author}{\bibfnamefont{P.~O.} \bibnamefont{Boykin}},
  \bibinfo{author}{\bibfnamefont{T.}~\bibnamefont{Mor}},
  \bibinfo{author}{\bibfnamefont{V.}~\bibnamefont{Roychowdhury}},
  \bibnamefont{and} \bibinfo{author}{\bibfnamefont{F.}~\bibnamefont{Vatan}}
  (\bibinfo{year}{1999}), \bibinfo{note}{quant-ph/9907067}.

\bibitem[{\citenamefont{Shor}(1996)}]{Shor_faults}
\bibinfo{author}{\bibfnamefont{P.~W.} \bibnamefont{Shor}}, in
  \emph{\bibinfo{booktitle}{37th Annual Symposium on Foundations of Computer
  Science}} (\bibinfo{publisher}{IEEE Press}, \bibinfo{year}{1996}), pp.
  \bibinfo{pages}{56--65}.

\bibitem[{\citenamefont{Szkopek et~al.}(2006)\citenamefont{Szkopek, Boykin,
  Fan, Roychowdhury, Yablonovitch, Simms, Gyure, and Fong}}]{boykin-2006}
\bibinfo{author}{\bibfnamefont{T.}~\bibnamefont{Szkopek}},
  \bibinfo{author}{\bibfnamefont{P.}~\bibnamefont{Boykin}},
  \bibinfo{author}{\bibfnamefont{H.}~\bibnamefont{Fan}},
  \bibinfo{author}{\bibfnamefont{V.}~\bibnamefont{Roychowdhury}},
  \bibinfo{author}{\bibfnamefont{E.}~\bibnamefont{Yablonovitch}},
  \bibinfo{author}{\bibfnamefont{G.}~\bibnamefont{Simms}},
  \bibinfo{author}{\bibfnamefont{M.}~\bibnamefont{Gyure}}, \bibnamefont{and}
  \bibinfo{author}{\bibfnamefont{B.}~\bibnamefont{Fong}},
  \bibinfo{journal}{IEEE Trans. Nano.} \textbf{\bibinfo{volume}{5}},
  \bibinfo{pages}{42} (\bibinfo{year}{2006}).

\bibitem[{\citenamefont{Knill}(2004)}]{Knill:04}
\bibinfo{author}{\bibfnamefont{E.}~\bibnamefont{Knill}},
  \emph{\bibinfo{title}{Quantum computing with very noisy devices}}
  (\bibinfo{year}{2004}), \bibinfo{note}{quant-ph/0410199}.

\end{thebibliography}

\end{document}